\documentclass[twocolumn,prb,amsmath,amssymb,floatfix,superscriptaddress]{revtex4-2} 
\usepackage{amsmath}
\usepackage{amssymb}
\usepackage{graphicx}
\usepackage{times}
\usepackage{bm}
\usepackage{color}
\usepackage{hyperref}
\usepackage{graphicx}
\usepackage{upgreek}
\usepackage{scalerel}
\usepackage{multirow}
 
\usepackage{pgffor}
\usepackage{pdfpages}
\usepackage{mathdots}
\usepackage{float}
\usepackage{silence}
\usepackage{physics} 
\usepackage{lscape}   
\usepackage{color, colortbl}
\usepackage[table]{xcolor}
\usepackage{comment}  
\makeatletter
\AtBeginDocument{\let\LS@rot\@undefined}
\makeatother

\newcommand{\red}[1] {\textcolor{black}{#1}}

\newcommand{\nn}{\nonumber}
\newcommand{\beq}{\begin{equation}}
\newcommand{\eeq}{\end{equation}}

\begin{document}

\title{Optical-vortex-pulse induced nonequilibrium spin textures in 
spin-orbit coupled electrons}

\author{Shunki Yamamoto}
\affiliation{Department of Materials Engineering Science, The University of Osaka, Toyonaka, Osaka 560-8531, Japan}
\author{Masahiro Sato}
\affiliation{Department of Physics, Chiba University, Chiba 263-8522, Japan}
\author{Satoshi Fujimoto}
\affiliation{Department of Materials Engineering Science, The University of Osaka, Toyonaka, Osaka 560-8531, Japan}
\affiliation{Center for Quantum Information and Quantum Biology, The University of Osaka, Toyonaka 560-8531, Japan}
\affiliation{Center for Spintronics Research Network, Graduate School of Engineering Science, The University of Osaka, Toyonaka, Osaka 560-8531, Japan}
\affiliation{Division of Spintronics Research Network, Institute for Open and Transdisciplinary Research Initiatives, The University of Osaka, Toyonaka, Osaka 560-8531, Japan}
\author{Takeshi Mizushima}
\affiliation{Department of Materials Engineering Science, The University of Osaka, Toyonaka, Osaka 560-8531, Japan}
\date{\today}

\begin{abstract}
Optical vortex beams are a type of topological light characterized by their inherent orbital angular momentum, leading to the propagation of a spiral-shaped wavefront. In this study, we focus on two-dimensional electrons with Rashba and Dresselhaus spin-orbit interactions and examine how they respond to pulsed vortex beams in the terahertz frequency band. Spin-orbital interactions play a vital role in transferring the orbital angular momentum of light to electron systems and generating spatiotemporal spin textures. We show that the spatiotemporal spin polarization of electrons reflects orbital angular momentum carried by optical vortex pulses. These findings demonstrate how optical vortices facilitate ultrafast spin manipulation in spin-orbit-coupled electrons. Our results can be straightforwardly extended to the case of higher-frequency vortex beams for other two-dimensional metals with a larger Fermi energy.
\end{abstract} 

\maketitle

\section{Introduction}

Vortices are prevalent phenomena that exist throughout nature, affecting both quantum and classical fluids. In 1992, Allen {\it et al.} introduced the concept of optical vortices~\cite{Allen_1992}, paving the way for a new connection between optics and quantum mechanics. These optical vortices maintain their topological charges, carrying an orbital angular momentum (OAM) of $m\hbar$ per photon ($m\in\mathbb{Z}$).
The unique features of optical vortices, such as their helical structure and doughnut-shaped intensity profile with phase singularities, have resulted in widespread applications. These applications include optical trapping and manipulation~\cite{he95,kug97,nei02,mac02,cur03,dho11}, optical tweezers~\cite{Kuga_1997,Neil_2002}, nanofabrication~\cite{oma10,toy12,toy13}, and optical communications~\cite{Barrerio_2008,Wang_2012,Bozinovic_2013}. In addition, advancements in laser technology enable the generation of optical vortices in the terahertz (THz) frequency range, which has garnered interest in condensed matter physics~\cite{qui22}. This is exemplified by their use in irradiating topological insulators~\cite{shintani_2016}, conducting electron systems~\cite{fuj18,fuj19,tak18}, 
superconductors~\cite{mizushima_2023,tod23,yer24,yeh24,yeh25,kang25}, magnetic materials~\cite{sato1_2017, sato2_2017,sir19,sir21,yav23}, 
and ferroelectric materials~\cite{gao23,gao24}.

Low-dimensional electron systems serve as an ideal platform for exploring emergent quantum phenomena~\cite{man15,sou16,bih22}. 
In fact, such low-dimensional systems can feel directly the electromagnetic field of applied vortex beams without screening the field. 
A key element is the intrinsic spin-momentum locking of electrons through spin-orbit interactions (SOIs)~\cite{dre55,byc84}. SOIs provide a foundation for electric and optical control of spins, such as current-induced spin polarization~\cite{aro89,edelstein_1990,aro91,mis04}, spin-charge conversion~\cite{sar04,and17,fen17}, and the photogalvanic effect~\cite{gan03,die07,moo10,mci12}, paving the way for spintronics applications. SOIs are also crucial for achieving topological superconductivity and are the heart of noncentrosymmetric superconductors~\cite{gor01,sat09-1,sat09-2,sat10,lut10,sau10,ali10,bau,yip14}.

Two-dimensional electron gases (2DEGs) in the interface of semiconductor-semiconductor heterostructures inherently exhibit SOIs due to the lack of inversion symmetry. There are two types of SOIs: Rashba~\cite{byc84} and Dresselhaus SOIs~\cite{dre55}. Rashba-type SOIs are realized in low-dimensional electrons confined at the quantum well, where inversion symmetry is explicitly broken. The broken inversion symmetry generates an electric field that interacts with the spin of itinerant electrons, leading to the splitting of their dispersion. Dresselhaus-type SOIs stem from the breaking of inversion symmetry in the underlying crystal. The SOIs also cause the spin splitting of the electron dispersions with their form depending on the growth direction relative to the crystallographic axis.
These SOIs, which are ubiquitous in low dimensions, facilitate interactions between electron spin and light, enabling optical manipulation of spin and serving as essential components for spintronic technology.

The combination of topologically structured light~\cite{zha09,she19,for19,nap23} 
and spin-orbit-coupled electrons allows for developing ultrafast and tailor-made manipulation of spins~\cite{cla13,sor19,ji20,Ishihara_2023,mat24,ter25}. 
Recently, it has been reported that vector vortex beams, which are structured light with spatially variant polarization, can manipulate the electron spin texture in the GaAs/AlGaAs quantum well by imprinting their spatial helicity structure ~\cite{Ishihara_2023}. 
Optical vortices are another type of topologically structured light characterized by OAM, yet their specific role in ultrafast spin manipulation remains unclear. Specifically, Laguerre-Gaussian beams having photon energy near the band gap of GaAs have been used to investigate photoinduced spin polarization in semiconductors~\cite{cla13}. However, light with OAM up to $\pm 5\hbar$ cannot detect the OAM dependence of spin polarization. This is because the OAM of light does not significantly change the optical selection rules for interband transitions when an azimuthal photon momentum is much smaller than the momentum scale of electrons. Recently, it has been demonstrated that the OAM of light can induce a nonlocal interaction between light and matter, resulting in the distinct OAM dependence of nonlocal photocurrent~\cite{ji20}.

In this paper, we examine the spin response of spin-orbit-coupled electron systems to optical vortices [Fig.~\ref{fig:setup}(a)]. In particular, we focus on the influence of excitations within the conduction bands \red{in the metallic states}. This situation differs from previous experiments~\cite{cla13,ji20} that stimulate optical interband transitions occurring beyond the band gap. 
We perform numerical calculations on the spin response using linear response theory, considering both the case involving only the Rashba SOI and the case of the equal strengths of the Rashba and Dresselhaus SOIs. \red{We theoretically analyze pulse lasers rather than continuous waves because many experiments use pulsed light fields to observe laser-driven spin textures.} Here we demonstrate that vortex beams can imprint characteristic spin textures reflecting the OAM of light~\cite{levitov_1985,edelstein_1990,fuj05,fujimoto_2007,Fujimoto2012}, when either Rashba or Dresselhaus SOI is dominant. We also show that the resultant spin response yields a clear dependence on the OAM of light. The necessary strength of the AC electric field, $E_0$, for detection is estimated to be $E_0=O(1{\rm -}10~{\rm kV}/{\rm cm})$, based on the realistic SOI values in GaAs/AlGaAs heterostructures. In our calculations, we consider vortex beams in the THz frequency range~\cite{miy19,ari20},
whose energy is much smaller than the band gap in semiconductor-semiconductor heterostructures. However, our results are applicable to optical vortices within the infrared and visible frequency ranges when the materials have a large Fermi surface and strong SOIs, such as heavy metals. 

\red{We note that when calculating laser-pulse-driven real-space spin textures in a spin-orbit-coupled metal, summing the optical transitions of each atom is not efficient. This is because, in metallic systems, many electrons with different wave vectors near the Fermi surface are excited simultaneously, which is quite different from the case of semiconductors. Specifically, in the former, the wave-vector space picture is effective for computing the laser response, whereas in the latter, direct calculation in real space is possible. Therefore, we will examine the laser-driven spin response by combining the linear response theory in wave-vector space with numerical calculations. We demonstrate that these microscopic analyses can produce laser-driven spin textures in metallic systems at a quantitative level.}

This paper is organized as follows. In Sec.~\ref{sec:model}, we present a model Hamiltonian of a 2DEG system with both Rashba and Dresselhaus SOIs and formulate the dynamical spin response to vortex beams using the linear response theory. We also introduce the Laguerre-Gaussian vortex beams. Numerical results on the spin response of 2DEGs to pulsed vortex beams are present in Sec.~III. We show that characteristic spin textures can be generated by vortex beams, depending on the sorts of SOIs. 
Section~IV is devoted to the conclusion and discussion. 
In this last section, we also mention an experimental method to observe the photo-induced spin textures.
Throughout this paper, we set $\hbar=k_{\rm B}=1$.

\section{Model and optical vortices}
\label{sec:model}

\subsection{Model Hamiltonian and symmetries}

\begin{figure}[t!]
    \includegraphics[width=8.5cm,clip]{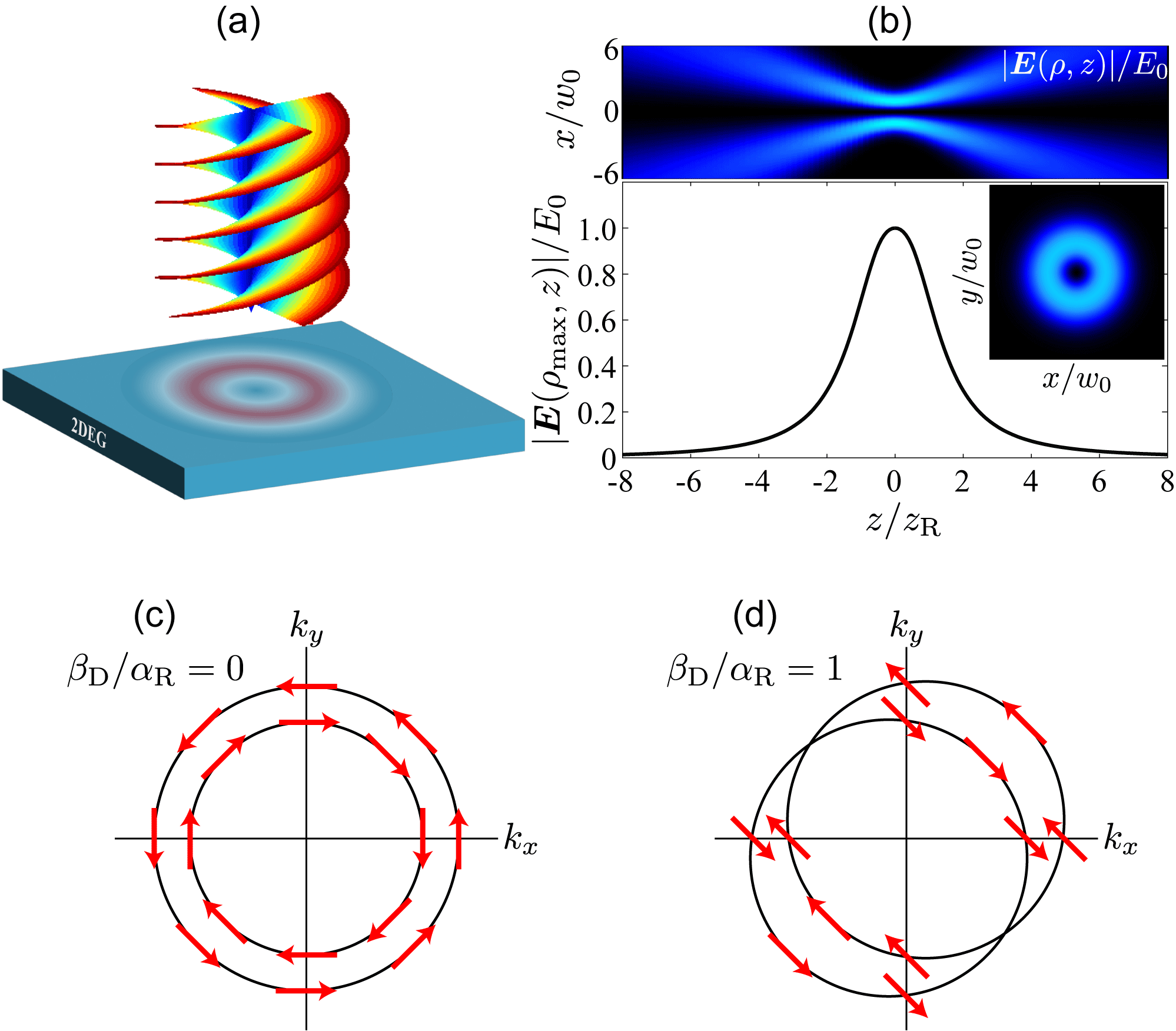}
    \caption{(a) Schematic image of our setup. A pulsed vortex beam is irradiated to a 2DEG formed in the quantum well of the GaAs/AlGaAs heterostructure. \red{(b) The spatial profile of the electric field generated by the vortex beam along the propagation ($z$) axis, $|{\bm E}(x,y=0,z)|$ (the upper panel) and $|{\bm E}(\rho_{\rm max},z)|$ (the lower pannel), where $(\rho,z)$ is the cylindrical coordinate and $\rho_{\rm max}$ is defined so that $E_0\equiv \max|{\bm E}(\rho,z=0)|$ at $\rho=\rho_{\rm max}$. The electric field sharply peaks within the Rayleigh range, $z_{\rm R}=O(100~\mu{\rm m})$. The inset is the electric field at the focal plane ($z=0$), where $x/w_0,~y/w_0\in [-3,3]$. Here we consider the vortex beam carrying the orbital angular momentum, $m=2$. For the further details, see Sec.~\ref{sec:vortex}.} (c,d) Fermi surfaces (thick curves) and spin polarization (arrows) of spin-orbit-coupled electrons on the Fermi surfaces: (c) $\beta_{\rm D}/\alpha_{\rm R}=0$ and (d) $\beta_{\rm D}/\alpha_{\rm R}=1$.}
    \label{fig:setup}
\end{figure}

In this paper, we examine the magnetic response of spin-orbit-coupled electron systems to irradiated vortex beams [Fig.~\ref{fig:setup}(a)]. \red{As shown in Fig.~\ref{fig:setup}(b), the intensity of the electric field associated with the vortex beam sharply peaks along its propagation axis within the Rayleigh range, $z_{\rm R}$, which is $O(100~\mu{\rm m})$ in the THz frequency range. For the further details, see Sec.~\ref{sec:vortex}.} Let us specifically consider the interface of the semiconductor-semiconductor GaAs/AlGaAs heterostructure, which is much thinner than the Rayleigh range. At the interface, the inversion symmetry is broken, allowing the Rashba SOI to be odd in the wave vector ${\bm k}$. We also consider another type of SOI that stems from the bulk inversion asymmetry of the underlying crystal, which is known as Dresselhaus SOI.

Let ${\bm \psi}^{\dag}_{\bm k} = (\psi^{\dag}_{\uparrow,{\bm k}}, \psi^{\dag}_{\downarrow,{\bm k}})$ be the spinor that consists of the creation operators of electrons with spin $\sigma=\uparrow,\downarrow$ and momentum ${\bm k}=(k_x,k_y)$, where $k_x$ and $k_y$ are taken along the [100] and
[010] directions, respectively. The Hamiltonian for 2DEGs with both Rashba and Dresselhaus SOIs is then given by
\begin{align}
\mathcal{H}
=\sum_{\bm k} {\bm \psi}^{\dag}_{\bm k}
H({\bm k})
{\bm \psi}_{\bm k}
=\sum_{\bm k} {\bm \psi}^{\dag}_{\bm k}
\left[ h_0({\bm k})+{\bm h}({\bm k})\cdot {\bm \sigma}\right]
{\bm \psi}_{\bm k},
\label{eq:H}
\end{align}
The Pauli matrices in the spin space are defined as ${\bm \sigma}=(\sigma_x,\sigma_y,\sigma_z)$. The scalar and vector components in Eq.~\eqref{eq:H} are given by 
\begin{gather}
h_0({\bm k}) = \frac{k^2}{2m_{\rm eff}} - \mu ,
\label{eq:H0} \\
{\bm h}({\bm k})=\alpha_{\rm R}\left(k_y\hat{\bm x}-k_x\hat{\bm y}\right) 
+ \beta_{\rm D}\left(k_x\hat{\bm x}-k_y\hat{\bm y}\right),
\label{eq:hk}
\end{gather}
where the unit vectors $\hat{\bm x}$ and $\hat{\bm y}$ are defined as $\hat{\bm x}=(1,0,0)$ and $\hat{\bm y}=(0,1,0)$.
Equation~\eqref{eq:H0} is the Hamiltonian density for free electrons with the effective mass, $m_{\rm eff}$, and the chemical potential, $\mu$. The vector components, ${\bm h}({\bm k})$, represent the effective in-plane magnetic field induced by the SOIs, where the strengths of the Rashba and Dresselhaus SOIs are denoted by $\alpha_{\rm R}$ and $\beta_{\rm D}$, respectively. 
The Hamiltonian can model electrons residing in III-V semiconductor quantum wells~\cite{koh12}. Electrons are confined to the two-dimensional quantum well formed at the interface of a semiconductor-semiconductor heterostructure. Here we take into account only excitations occurring within the conduction band, disregarding excitations from the valence band beyond the band gap. The latter excitations are insignificant when the photon energy of optical vortices is considerably lower than the band gap.

The energy dispersions of the conduction electrons are obtained by diagonalizing the Hamiltonian whose eigenenergy is 
\beq
E_{s}({\bm k}) = h_0 ({\bm k}) + s|{\bm h}({\bm k})|,
\label{eq:E0}
\eeq
where $s=\pm$ is the band index. In Figs.~\ref{fig:setup}(b) and \ref{fig:setup}(c), we show the Fermi surfaces and ${\bm h}({\bm k})$ obtained from the model Hamiltonian for $\beta_{\rm D}/\alpha_{\rm R}=0$ and $\beta_{\rm D}/\alpha_{\rm R}=1$, respectively. We also define the Fermi energy and the Fermi wave number of electrons without SOIs as $\varepsilon_{\rm F}$ and $k_{\rm F}\equiv \sqrt{2m_{\rm eff}\varepsilon_{\rm F}/\hbar^2}$, respectively. 
The arrows in Figs.~\ref{fig:setup}(b) and \ref{fig:setup}(c) denote the spin polarization direction of electrons on the Fermi surfaces and they indicate the spin-momentum locking.

It is important to summarize the symmetries that the Hamiltonian in Eqs.~\eqref{eq:H0} and \eqref{eq:hk} holds. In the case of purely Rashba SOIs ($\beta _{\rm D}=0$), the Hamiltonian holds the $C_{4v}$ symmetry about the $z$-axis as
\begin{align}
U_4{H}(k_x,k_y)U_4^{\dag} = H(k_y,-k_x), 
\label{eq:C4}
\end{align}
where $U_4\equiv e^{-i(\pi/4)\sigma_z}$ denotes the fourfold rotation about the $z$ axis. At $\alpha_{\rm R}=\beta_{\rm D}$, the Hamiltonian can be transformed into the diagonal form by jointly rotating the spatial coordinate and the spin space as~\cite{ber06} 
\begin{align}
\tilde{H}(k_+,k_-) \equiv &
UH(R^{-1}_2{\bm k})U^{\dag}  \nn \\
=& \frac{k_+^2+k^2_-}{2m_{\rm eff}}-\mu - 2\alpha_{\rm R}k_+{\sigma}_z,
\label{eq:HSU2}
\end{align}
where $R_2$ is the rotation matrix about the $z$-axis by $\pi/4$ and we set $k_{\pm}=(k_y\pm k_x)/\sqrt{2}$. The ${\rm SU}(2)$ matrix, $U$, denotes the global spin rotation about the $(\hat{\bm x}+\hat{\bm y})/\sqrt{2}$ direction by $\pi/2$ 
\begin{align}
U = \frac{1}{\sqrt{2}}\left[
\sigma_0 -\frac{i}{\sqrt{2}}(\sigma_x+\sigma_y)
\right],
\label{eq:SU2}
\end{align}
where $\sigma_0$ is the unit matrix in the spin space. 
The Hamiltonian at $\alpha_{\rm R}=\beta_{\rm D}$ remains invariant under the exact ${\rm SU}(2)$ symmetry~\cite{ber06}. Thus, we below refer to this type of SOI as the ${\rm SU}(2)$ symmetric SOI.

Nonmagnetic impurity effects are taken into account as spin-independent point-like scattering centers, 
$u_{\rm imp}({\bm r}) = \sum_{i} u_0 \delta({\bm r}-{\bm R}_i)$, 
where ${\bm R}_i$ is the position of an impurity atom and $u_0$ is the impurity potential. The impurity centers are randomly distributed in the two-dimensional plane. The first moment vanishes, $\langle u_{\rm imp}({\bm r})\rangle_{\rm imp} = 0$ and the second moment is finite, $\langle |u_{\rm imp}({\bm k},{\bm k}^{\prime})|^2\rangle_{\rm imp} = n_{\rm imp}u^2_0$, where $n_{\rm imp}$ is the impurity concentration. Let $G({\bm k},i\varepsilon_n)$ be the impurity averaged Green's function of the 2DEGs in Matsubara representation. In the clean limit, the $2\times 2$ spin matrix form of the Green's function reduces to $G({\bm k},i\varepsilon_n)\equiv [i\varepsilon_n-H({\bm k})]^{-1}$, where $\varepsilon_n \equiv (2n+1)\pi T$ is the fermionic Matsubara frequency at temperature $T$ ($n\in\mathbb{Z}$). On the other hand, in the case of a finite impurity density, the Green's function is expressed in the Born approximation as 
\begin{gather}
    G({\bm k},i\varepsilon_n)
    = \sum_{s=\pm} \frac{1+s\hat{\bm h}({\bm k})\cdot{\bm \sigma}}{2}G_s({\bm k},i\varepsilon_n), \label{eq:G} \\
    G_s({\bm k},i\varepsilon_n) = \frac{1}{i\varepsilon_n - E_s({\bm k}) + i\Gamma}.
    \label{eq:Gs}
\end{gather}
The relaxation rate $\Gamma=-{\rm Im}\Sigma$ is related to the imaginary part of the impurity self-energy ($\Sigma$), which is obtained in the Born approximation as
\beq
\Sigma(i\varepsilon_n) =  n_{\rm imp}u^2_0 \int \frac{d^2{\bm k}}{(2\pi)^2}
G({\bm k},i\varepsilon_n).
\label{eq:sigma}
\eeq
The imaginary part of the self-energy, $\Gamma$, is related to the relaxation time, $\tau$, as $\Gamma = \frac{\hbar}{2\tau}{\rm sgn}(\varepsilon_n)$.
In this work, we consider a metallic state, that is, $\Gamma \ll \varepsilon_{\rm F}$.

\subsection{Spin response of 2DEGs}
\label{sec:linear}

Here we formulate the spin response of 2DEGs to vortex beams. Let ${\bm E}$ be the electric field associated with the vortex beams. In the linear response regime on ${\bm E}$, the spin response of the 2DEGs is obtained as 
\begin{align}
    S_{\mu}({\bm x},t)=\int dt^{\prime}\int d{\bm x}^{\prime}\Upsilon_{\mu\nu}(\bm{x}-{\bm x}^{\prime},t-t^{\prime})E_{\nu}(\bm{x}^{\prime},t^{\prime}).
\end{align}
The repeated Greek indices imply the sum over $x,y,z$. The Fourier component of the function, $\Upsilon$, is obtained as $\Upsilon_{\mu\nu}\equiv \chi_{\mu\nu}/i\omega$ from the spin-current correlation functions, $\chi_{\mu\nu}$. The spin-current correlation functions are written in the Matsubara representation as
\begin{align}
\chi_{\mu\nu}(q)
&\equiv -\int d\tau
\left\langle 
\hat{S}_{\mu}({\bm q},\tau)\hat{j}_{\nu}(-{\bm q},\tau)
\right\rangle
e^{i\omega_m\tau}\nn \\
&=\sum_{k}\mathrm{tr}_2[G({k})\hat{S}_{\mu}G({k}+{q})\hat{j}_{\nu}(k)] 
\label{eq:jscf}.
\end{align}
where $\tau$ is the imaginary time and the symbol $\mathrm{tr}_{2}$ means trace in the $2\times 2$ spin space. We also introduced the abbreviations, $k\equiv ({\bm k},i\varepsilon_n)$, ${q}\equiv ({\bm q},i\omega_m)$, and $\sum_{k} \equiv \int \frac{d^2{\bm k}}{(2\pi)^2}T\sum_n$. The Matsubara frequency for bosons at temperature $T$ is defined as $\omega_m=(2m+1)\pi T$, where $m\in\mathbb{Z}$. 
The spin and current operators are given by
$\hat{S}_{\mu}=\frac{\hbar}{2}\sigma_{\mu}$ and 
$\hat{j}_{\nu}({\bm k})=-e(\partial H({\bm k})/\partial k_{\nu})$, respectively.

Let us define the retarded and advanced Green's functions as 
$G^{\rm R}({\bm k},\varepsilon)\equiv G({\bm k},i\varepsilon_n\rightarrow \varepsilon + i0_+)$ and 
$G^{\rm A}({\bm k},\varepsilon)\equiv G({\bm k},i\varepsilon_n\rightarrow \varepsilon - i0_+)$, where $0_+$ is an infinitesimal positive constant. Performing the Matsubara sum in Eq.~\eqref{eq:jscf}, the correlation functions reduce to 
\begin{align}
\chi_{\mu\nu}(q) =& -\int\frac{d^2{\bm k}}{(2\pi)^2}\int \frac{d\varepsilon}{2\pi}f(\varepsilon){\rm tr}_2
\Gamma_{\mu\nu}(k,q),
\end{align}
where $f(\varepsilon)=1/(e^{\varepsilon/T}+1)$ is the Fermi distribution function and ${\rm tr}_2$ denotes the trace over the spin space. We have introduced the shorthanded notation, ${k}\equiv ({\bm k},\varepsilon)$ and ${q}\equiv ({\bm q},\omega)$. The matrices, $\Gamma_{\mu\nu}({k},{q})$, are defined as 
\begin{align}
\Gamma_{\mu\nu}({k},{q})
=& \left\{G^{\rm R}({k})-
G^{\rm A}({k})
\right\}\hat{\Lambda}_{\mu}G^{\rm R}({k}+{q})j_{\nu}({\bm k}) \nn \\
&+ G^{\rm A}({k}-{q})\hat{\Lambda}_{\mu}\left\{
G^{\rm R}({k})-G^{\rm A}({k})
\right\}j_{\nu}({\bm k}).
\label{eq:gamma1}
\end{align}
In Eq.~\eqref{eq:gamma1}, we have replaced the bare spin vertex function $\hat{S}_{\mu}$ to the renormalized spin vertex function $\hat{\Lambda}_{\mu}$, which incorporates the sum of the usual ladder diagrams of impurity scattering (see Fig.~\ref{fig:diagram}). In the current situation, the frequency of the irradiated beam is in the THz band, $\Omega \sim \varepsilon_{\rm F} = O(10~{\rm meV})$, while the scattering rate of two-dimensional electrons in AlGaAs/GaAs heterostructures is estimated as $\Gamma \approx 0.1 (k_{\rm F}\alpha_{\rm R}) \ll \varepsilon_{\rm F}$ (see Sec.~\ref{sec:setup}). This situation satisfies the weak impurity condition,  $\Omega \tau \gg 1$, where $\tau \equiv \hbar /2\Gamma$ is the relaxation time. For $\Omega \tau \gg 1$, the impurity vertex function in Eq.\eqref{eq:vc} reduces to the bare vertex function, $\hat{\Lambda}_{\mu}\approx\hat{S}_{\mu}$ for all $\beta_{\rm D}/\alpha_{\rm R}$. Therefore, we disregard the effect of the vertex corrections on the spin response in the subsequent calculations. The details are described in Appendix~\ref{sec:vertex}.

\subsection{Vortex beams}
\label{sec:vortex}

Vortex beams, which are also referred to as Laguerre-Gaussian beams, are a type of topological light characterized by OAM and spin angular momentum (SAM)~\cite{Allen_1992,hall_1996}.
Let us first consider a monochromatic vortex beam propagating along the $z$-direction. The vortex beam with OAM ($m \in\mathbb{Z}$) and SAM ($\lambda=0,\pm 1$) generates the electric field, which in a monochromatic light of frequency $\Omega$ is described as
\begin{align}
    &\bm{E}
    ({\bm x},t)=
    \frac{1}{2}\left[
    u_{p,m}(\boldsymbol{x})e^{-i\Omega t}\hat{\boldsymbol{e}}_{\lambda} 
    +u_{p,m}^{\ast}(\boldsymbol{x})e^{i\Omega t}\hat{\boldsymbol{e}}^{\ast}_{\lambda}
    \right].
    \label{eq:E} 
\end{align}
Since the wavelength of THz vortex beams is significantly longer than the thickness of the two-dimensional quantum well at the interface, we can disregard the $z$-dependence of ${\bm E}$. 
The polarization vector is defined as $\hat{\bm e}_{\lambda}=\lambda(\hat{\bm x}+i\lambda \hat{\bm y})/\sqrt{2}$ with the SAM of light, $\lambda$, where $\lambda=\pm 1$ ($\lambda=0$) denote circularly polarized (linearly polarized) light. 

\red{The spatial profile of $u_{p,m}$ for vortex beams is determined by solving the Helmholtz equation, $({\bm \nabla}^2+k^2)u_{p,m}=0$, in the cylindrical coordinate, where $k\equiv \Omega/c$ with the speed of light, $c$. To solve the equation, we employ the paraxial approximation that assumes $u_{p,m}({\bm x})=\tilde{u}_{p,m}({\bm x})e^{ikz}$ to satisfy the conditions, $| \frac{\partial^2\tilde{u}}{\partial z^2}|\ll | \frac{\partial^2\tilde{u}}{\partial x^2}|$, $|\frac{\partial^2\tilde{u}}{\partial y^2}|$, and $| \frac{\partial^2\tilde{u}}{\partial z^2}|\ll 2 k| \frac{\partial \tilde{u}}{\partial z}|$. The amplitude distribution changes slowly with distance $z$ when compared to variations of $\tilde{u}({\bm x})$ in the lateral direction. Then, 
we cacn remove the second-order derivative term and the Helmholtz equation reduces to  
\beq
\left( {\bm \nabla}^2_{\perp} + 2ik\frac{\partial}{\partial z} \right)\tilde{u}({\bm x})=0,
\label{eq:Helm}
\eeq
where $\Delta_{\perp}^2\equiv \partial^2_x+\partial^2_y$. The Laguerre-Gaussian modes form a complete set of this equation, which are given by
\begin{align}
u_{p,m}({\bm x}) =& \frac{1}{\sqrt{w(z)}}\left(\frac{\sqrt{2}\rho}{w(z)}\right)^{|m|}e^{-\frac{\rho^2}{w(z)^2}}L^{|m|}_p\left( 
\frac{2\rho^2}{w(z)^2}
\right)e^{im\theta}\nn \\
&\times \exp\left(-\frac{ik\rho^2z}{z^2+z^2_{\rm R}}-i(2p+m+1)\chi(z)\right),
\label{eq:uLG}
\end{align}
where $L^{m}_p$ is the associated Laguerre polynomial. We have introduced the cylindrical coordinate with $\rho=\sqrt{x^2+y^2}$ and $\theta = \tan^{-1}(y/x)$. The beam waist $w_0$ determines the width, $w(z)=w_0\sqrt{1+z^2/z^2_{\rm R}}$, and the Rayleigh range of the beam, $z_{\rm R} = kw^2_0/2$ with $k=\Omega/c$, where $c$ is the speed of light. The Rayleigh range is the distance along the $z$ axis from the focal plane ($z=0$) at which the cross section of the beam becomes twice its minimum value, indicating the length scale of sharply peaked intensities along the $z$ axis [see Fig.~\ref{fig:setup}(b)]. 
We have also introduced the Gouy phase $\chi(z)=\tan^{-1}(z/z_{\rm R})$.} 
At the focal plane ($z=0$), the LG mode reduces to 
\begin{align}
u_{p,m}(\rho,\theta) =& \frac{1}{\sqrt{w_0}}\left(\frac{\sqrt{2}\rho}{w_0}\right)^{|m|}e^{-\frac{\rho^2}{w^2_0}}L^{|m|}_p\left( 
\frac{2\rho^2}{w^2_0}
\right)e^{im\theta}.
\end{align}

The wavefront of the vortex beams, or the isophase plane, takes a spiral shape around the propagation ($z$) axis. \red{In Figs.~\ref{fig:CCW}(b1-b3) and \ref{fig:CCW}(d1-d3), we plot the intensity and the direction of the electric field induced by the vortex beams. For $m=0$ the beam reduces to the usual Gaussian beam whose intensity peaks at the center ($\rho=0$) [Fig.~\ref{fig:CCW}(b1)].} The single-valuedness of the electromagnetic fields requires $|u_{p,m}({\bm x})|$ to be zero at the center of the beam, resulting in a doughnut-shaped intensity profile in the transverse plane for $m\neq 0$. In momentum space, $|u_{p,m}({\bm q})|$, the intensity of vortex beams also shows a doughnut shape and peaks around $|{\bm q}|\sim 1/w_0$. The radius $|{\bm q}|$ increases with $|m|$, indicating that for vortex beams with $m\neq 0$, nonzero $|{\bm q}|$ contributes to the spin-current response function. 

In subsequent sections, we also consider a pulsed vortex beam. The electric field generated by the pulsed vortex beam with $(m,s)$ is given by
\begin{align}
{\bm E}({\bm x},t) = &{\rm Re}\bigg\{\frac{E_0u_{p,m}({\bm x})}{\max | u_{p,m}({\bm x})|} \nn \\
& \times
\exp\left[
-\left( \frac{t-t_{0}}{\sigma_E}\right)^2 -i\Omega t
\right]\hat{\bm e}_s\bigg\},
\label{eq:pulse}
\end{align}
where $\Omega$ and $\sigma_E\equiv 2\pi n_{\rm p}/\Omega$ are the frequency and the full width at half-maximum of the beam intensity, respectively. In this work, we focus on vortex beams with $p=0$, i.e., single-doughnut types of the vortex beam.

\section{Numerical results}

\subsection{Calculated systems and parameters}
\label{sec:setup}

In this paper, we consider 2DEGs residing in semiconductor-semiconductor heterostructures. In such heterostructures, the strengths of the SOIs, $k_{\rm F}\alpha_{\rm R}$ and $k_{\rm F}\beta_{\rm D}$, are of the order of $1~{\rm meV}$ and less than the Fermi energy $\varepsilon_{\rm F}$. 
In the following numerical calculations, we set $k_{\rm F}\alpha = 0.1~{\rm meV}$~\cite{mil03,yam21,yam23}, which corresponds to the parameters of the GaAs/AlGaAs heterostructures, $\varepsilon_{\rm F}=20~{\rm meV}$ and $m_{\rm eff}=0.067m_{\rm e}$~\cite{mas85}, where $m_{\rm e}$ is the free electron mass. The set of these parameters satisfies $k_{\rm F}\alpha_{\rm R}\ll \varepsilon_{\rm F}$.
In Sec.~\ref{sec:rashba}, we focus on the spin response of 2DEGs with purely Rashba-type SOI ($\beta_{\rm D}/\alpha_{\rm R}=0$). The spin response in the case of ${\rm SU}(2)$ symmetric SOI with $\alpha_{\rm R}=\beta_{\rm D}$ is discussed in Sec.~\ref{sec:su2}. 
We also fix $T/\varepsilon_{\rm F}=0.1$ and $\Gamma=0.1(\alpha_{\rm R} k_{\rm F})$ in the following calculations, where the latter satisfies the weak impurity concentration condition, $\Omega\tau = \hbar\Omega/2\Gamma \gg 1$. \red{We note that the frequency spectrum of the pulsed field in Eq.~\eqref{eq:pulse} is sharply localized around $\Omega$ and the standard deviation $\sqrt{2}/\sigma_E$ is much smaller than $\Omega$ with the current parameter set. Therefore, the full frequency spectrum satisfies the weak impurity concentration.} We would also like to emphasize that the qualitative results are insensitive to the choice of these values. 

\begin{figure*}[t!]
    \includegraphics[width=15cm]{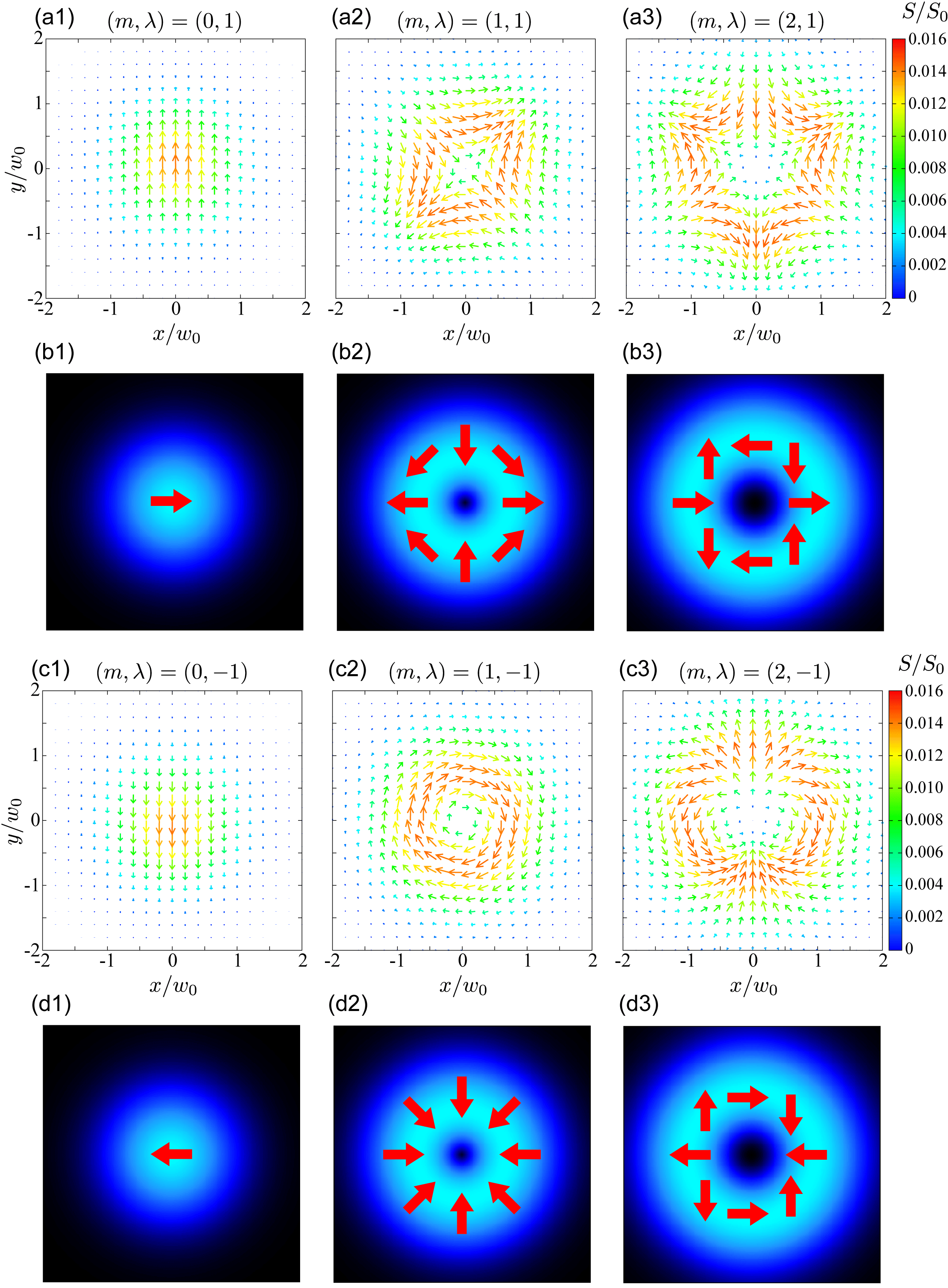}
    \caption{Linear response of the spin density, ${\bm S}({\bm x},t)/S_0$, induced by monochromatic vortex beams with $(\lambda,m)=(1,0)$ (a1), $(1,1)$ (a2), $(1,2)$ (a3), $(-1,0)$ (c1), $(-1,1)$ (c2), and $(-1,2)$ (c3), where $\lambda$ and $m$ are the SAM and OAM of light, respectively. The scaled quantity, $S({\bm x},t)/S_0$, with $S_0\equiv \hbar /2$ represents the spin density per $1~\mu{\rm m}^2$. We also plot the intensity of the electric field induced by the vortex beams with $(\lambda,m)=(1,0)$ (b1), $(1,1)$ (b2), $(1,2)$ (b3), $(-1,0)$ (d1), $(-1,1)$ (d2), and $(-1,2)$ (d3). Here we set $\alpha_{\rm D}k_{\rm F}=0.1~{\rm meV}$ and $\beta_{\rm D}=0$. The arrows in (a*) and (c*) correspond to the local spins $(S_x,S_y)/S_0$, and the color represents the amplitude of the local spin, $|{\bm S}(x,y)|/S_0$. The thick arrows in (b*) and (d*) show the direction of the electric field, ${\bm E}({\bm x})/|{\bm E}({\bm x})|$.}
    \label{fig:CCW}
\end{figure*}

A pulsed vortex beam is irradiated to the 2DEGs from $t=0$. In the numerical calculations, we take the number of cycles of the pulse field, the beam waist, and the maximum intensity of the electric field as $n_{\rm p}=3$, $w_0=2000 k^{-1}_{\rm F}$, $eE_0/\hbar \Omega k_{\rm F}=0.5$, respectively. Using $\Omega = 0.1\varepsilon_{\rm F}$, $\varepsilon_{\rm F}=20~{\rm meV}$, and $k_{\rm F}=0.1~{\rm nm}^{-1}$, these paraemeters are approximately estimated as $w_0 \sim 0.02~{\rm mm}$ and $E_0 \sim 0.1 {\rm kV}/{\rm cm}$. We also set $t_0=2\sigma_E$ and the size of the $xy$ plane to $x,y\in[-2w_0,2w_0]$. In the following, we present the numerical results of the spin density on the scale of $S_0\equiv \hbar /2$. The scaled quantity, $S({\bm x},t)/S_0$, represents the spin density per $1~\mu{\rm m}^2$ induced by an electric field intensity, $\max|{\bm E}({\bm x},t)|=0.1~{\rm kV}/{\rm cm}$.

Here we consider optical vortices with a frequency $\Omega \sim \varepsilon_{\rm F} = O(10~{\rm meV})$, corresponding to the THz frequency band. However, by considering heavy metals rather than semiconductor heterostructures, our results can also be applied to optical vortices within the infrared and visible frequency ranges.

\subsection{Rashba SOI: $\beta_{\rm D}/\alpha_{\rm R}=0$}
\label{sec:rashba}

Let us first consider the spin response of 2DEGs with Rashba-type SOIs, $\alpha_{\rm R}\neq 0$ and $\beta_{\rm D}=0$. Figure~\ref{fig:CCW} shows the linear response of the spin densities, $(S_x,S_y)$, to monochromatic vortex beams with the OAM, $m=0,1,2$, and the SAM, $\lambda=\pm 1$. We note that in the present model of 2DEGs, the longitudinal component of the spin does not respond to vortex beams, $S_z=0$. It is seen from Figs.~\ref{fig:CCW}(a) and \ref{fig:CCW}(c) that electron spins are highly excited in areas with strong electric field strength. For the $m=0$ vortex beam, corresponding to a Gaussian beam, the excited spins are uniformly aligned within the region, $\sqrt{x^2+y^2}\lesssim w_0 = 2000 k^{-1}_{\rm F}$, where the intensity of the electric field maximizes. In contrast, $m\neq 0$ vortex beams excite the spins along the circumference at the radius $\sim w_0 = 2000\lambda_{\rm F}$ where the intensity of the doughnut-shaped vortex beam peaks.  
When traveling around the circumference, the spins exhibit a characteristic texture in which their orientation gradually rotates. For the $m=1$ ($m=2$) vortex beam, the local spin rotates once (twice) clockwise along the doughnut-shaped region.

These results are understandable with the symmetry of the spin-current correlation functions in Eq.~\eqref{eq:jscf}. As shown in Eq.~\eqref{eq:C4}, the Hamiltonian for 2DEGs holds the $C_{4v}$ symmetry about the $z$-axis perpendicular to the plane. This symmetry leads to $\chi_{xx}=\chi_{yy}=0$ and $\chi_{xy}=-\chi_{yz}$ in $q\rightarrow 0$, implying that the local spin response points to the direction, 
\begin{align}
{\bm S}({\bm x},t) \propto \hat{\bm z}\times {\bm E}({\bm x},t).
\label{eq:S1}
\end{align}
For a monochromatic vortex beam, the profile of the local electric field generated by the vortex beam reads
\begin{align}
{\bm E}({\bm x},t) \propto  
\lambda\cos(m\theta-\omega t) \hat{\bm x}
- \sin(m\theta-\omega t) \hat{\bm y},
\end{align}
where $\theta$ is the azimuthal angle in the $xy$ plane. In Figs.~\ref{fig:CCW}(b) and \ref{fig:CCW}(d), we plot the local electric field of Eq.~\eqref{eq:E} induced by the vortex beams with $(\lambda,m)$ at $t=0$. It is seen from Figs.~\ref{fig:CCW}(a) and \ref{fig:CCW}(b) that the induced spins always orient perpendicular to the local electric field, as shown in Eq.~\eqref{eq:S1}, ${\bm S}({\bm x},t)\perp {\bm E}({\bm x},t)$. 
It is also seen from Eq.~\eqref{eq:hk} that the 2DEGs experience the local electric field ${\bm E}$ through the Rashba-type SOI term as 
${\bm h}({\bm k}-e{\bm A})\cdot{\bm \sigma} = {\bm h}({\bm k})\cdot{\bm \sigma}
+C_{\rm R}\left( {\bm E}(q)\times {\bm \sigma}\right) $,
where $C_{\rm R}\equiv e\alpha _{\rm R}/i\Omega$. Electron spins interact with an applied electric field that is perpendicular to their spin orientation in the momentum space. As mentioned in Sec.~\ref{sec:linear}, impurity vertex corrections do not affect their orientation for $\Omega \tau\gg 1$. Therefore, in the case of purely Rashba-type SOIs ($\beta_{\rm D}/\alpha_{\rm R}=0$), the local spin response of 2DEGs is transverse to the local electric field generated by vortex beams. 

\begin{figure}[t!]
    \includegraphics[width=85mm]{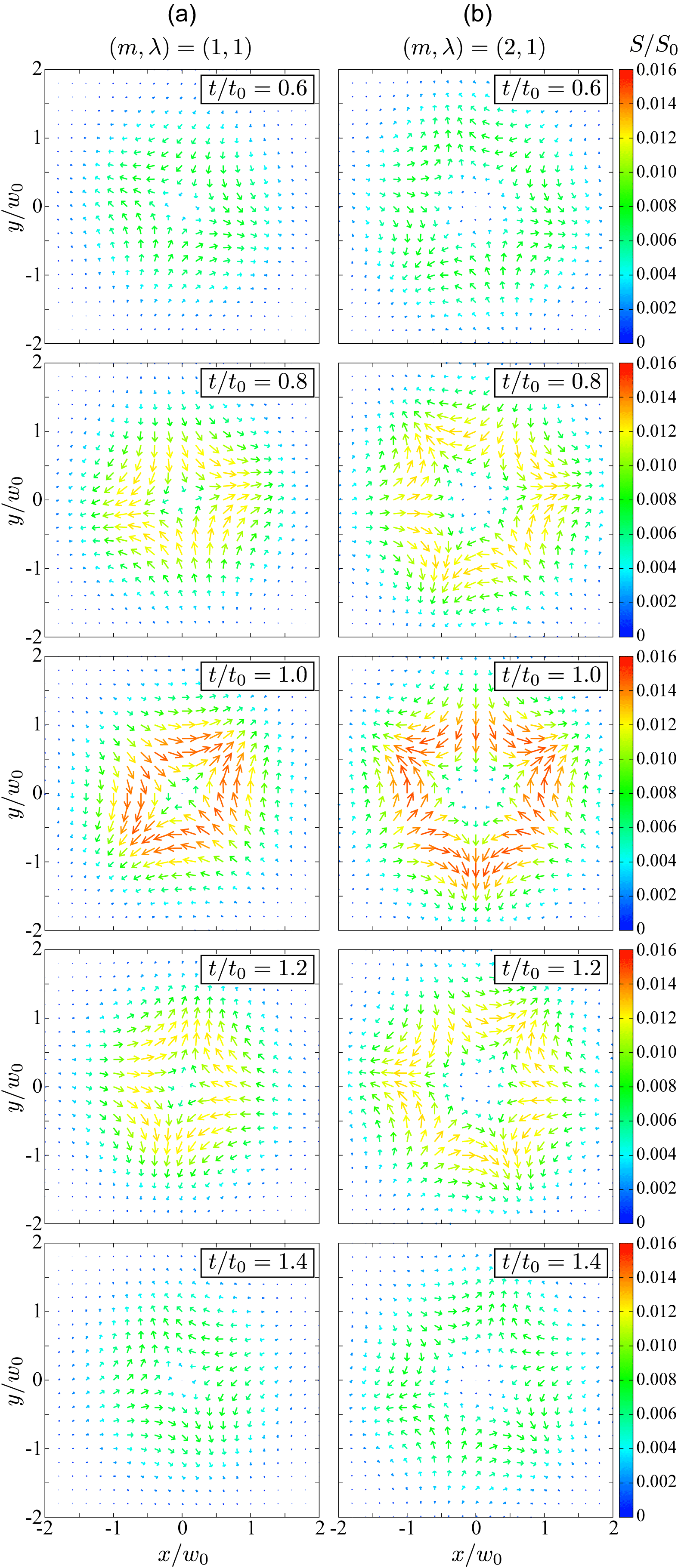}
    \caption{Snapshots of the local spin density in 2DEGs with $\alpha_{\rm R}k_{\rm F}= 0.1~{\rm meV}$ and $\beta_{\rm D}=0$ after the optical vortex pulse is irradiated. The OAM of light is set to be $m=1$ (a) and $m=2$ (b), where the SAM of light is fixed to $\lambda = 1$.}
    \label{pulse_RHC}
\end{figure}

The imprinted spin textures reflect the OAM carried by the applied electric field. The topological charge of the vortex beam with the OAM ($m$) and the SAM ($\lambda$) is defined by the winding number~\cite{shintani_2016}, 
\beq
w[{\bm E}] = \frac{1}{2\pi}\oint_C \epsilon _{\mu\nu}\hat{n}_{\mu}({\bm x},t) \partial _{x_j}\hat{n}_{\nu}({\bm x},t)dx_j
= - \lambda m,
\label{eq:w}
\eeq
where $\hat{\bm n}({\bm x},t)={\bm E}({\bm x},t)/|{\bm E}({\bm x},t)|$ is the unit vector pointing to the local electric field and $C$ is a closed path enclosing the singularity of the vortex beam at ${\bm x}={\bm 0}$. As shown in Figs.~\ref{fig:CCW}(a2) and \ref{fig:CCW}(a3), the local spins rotate in a clockwise direction once and twice, respectively, when moving counterclockwise around a circle of radius $w_0$. In a similar manner, the spins in Figs.~\ref{fig:CCW}(c2) and \ref{fig:CCW}(c3) rotate counterclockwise once and twice, respectively. The winding number of the imprinted spin textures is given by substituting $\hat{\bm n}=\hat{\bm S}/|\hat{\bm S}|$ in Eq.~\eqref{eq:w}, which coincides with that of the vortex beams, i.e., $w[{\bm S}]=w[{\bm E}]=-\lambda m$. This demonstrates that the OAM of light can be encoded into the spin textures of electrons through the SOI. This conclusion is also applicable to the case of purely Dresselhaus-type SOIs.

Figure~\ref{pulse_RHC} shows the time evolution of spin polarization following the irradiation of a pulsed vortex beam in Eq.~\eqref{eq:pulse}. The movies demonstrating the real-time evolution of local spins in 2DEGs irradiated by pulsed vortex beams are also available in Supplemental Materilal~\cite{SM}, where the SOI strengths are set to $\alpha_{\rm R}k_{\rm F}=0.1~{\rm meV}$ and $\beta_{\rm D}=0$. 
It can be seen that the spatial profile of the spin texture yields a vortex shape at all times. Similar to monochromatic vortex beams, the imprinted spin orientation reflects the OAM of light, $m$. Therefore, the OAM of vortex beams can be transferred to 2DEGs through the Rashba-type SOC, enabling the ultrafast manipulation of the spin texture. In contrast, conventional Gaussian beams with $m=0$ cause a uniform alignment of the spin orientation at all times.

So far, we have demonstrated that circularly polarized vortex beams can imprint the OAM of light onto the spin texture in 2DEGs with the Rashba SOI. The polarization of light plays an essential role in manipulating the spin texture. As shown in Eq.~\eqref{eq:S1}, the orientation of the local spins reflects the spatial profiles of the electric field generated by vortex beams. For linearly polarized light ($\lambda =0$), the electric field is aligned to the $\pm \hat{\bm y}$ directions, leading to the formation of spin domains alternatively oriented in the $+\hat{\bm x}$ and $-\hat{\bm x}$ directions.

\begin{figure}[t!]
    \includegraphics[width=85mm]{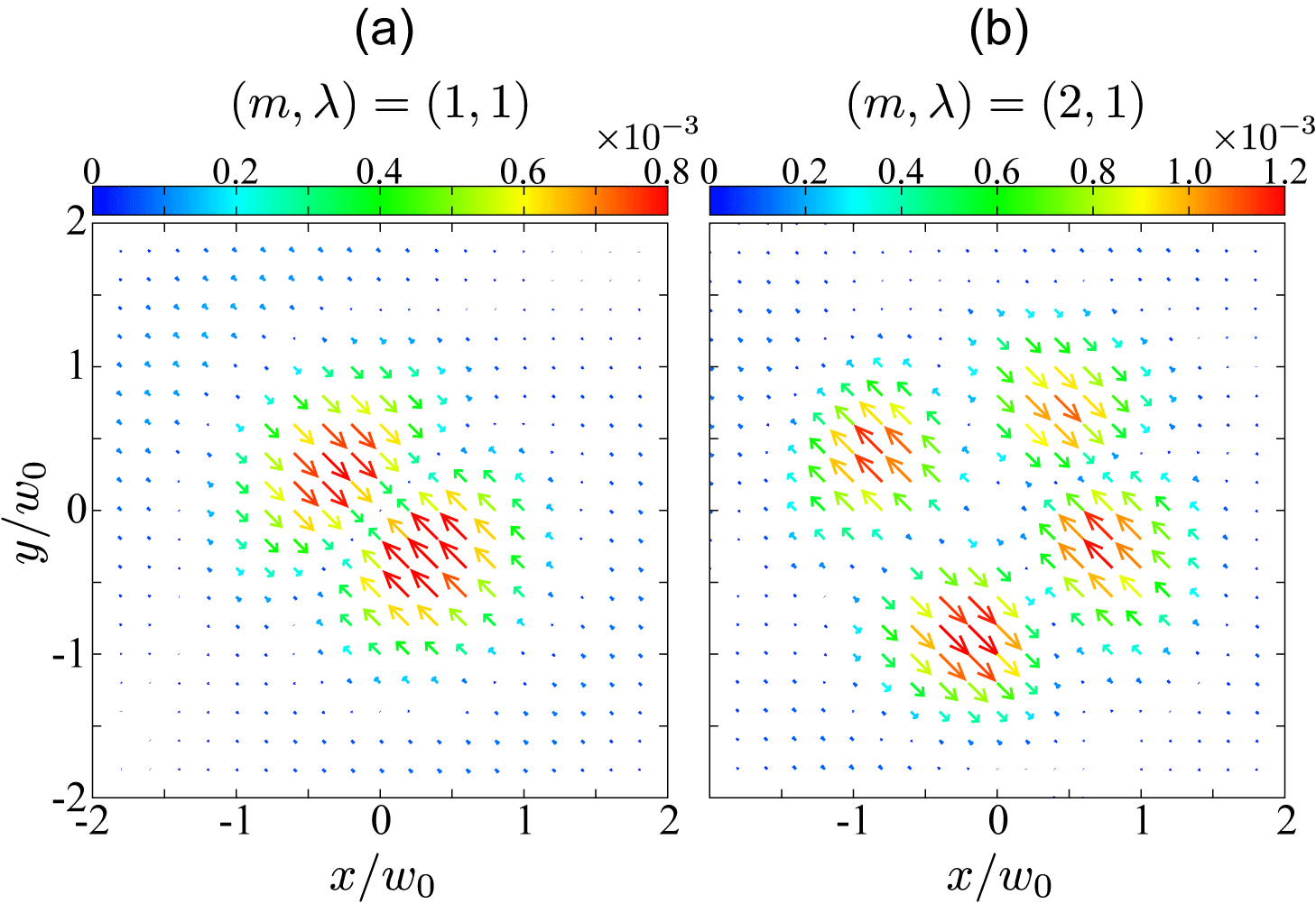}
    \caption{Local spin density in 2DEGs with the ${\rm SU}(2)$ symmetric SOI, $\alpha_{\rm R}k_{\rm F}=\beta_{\rm D}k_{\rm F}=0.1~{\rm meV}$ after the vortex beams with $(m,\lambda)=(1,1)$ and $(2,1)$ are irradiated. 
    }
    \label{fig:SU2-1}
\end{figure}

\subsection{${\rm SU}(2)$ symmetric SOI: $\beta_{\rm D}=\alpha_{\rm R}$}
\label{sec:su2}

\begin{figure}[t!]
    \includegraphics[width=82mm]{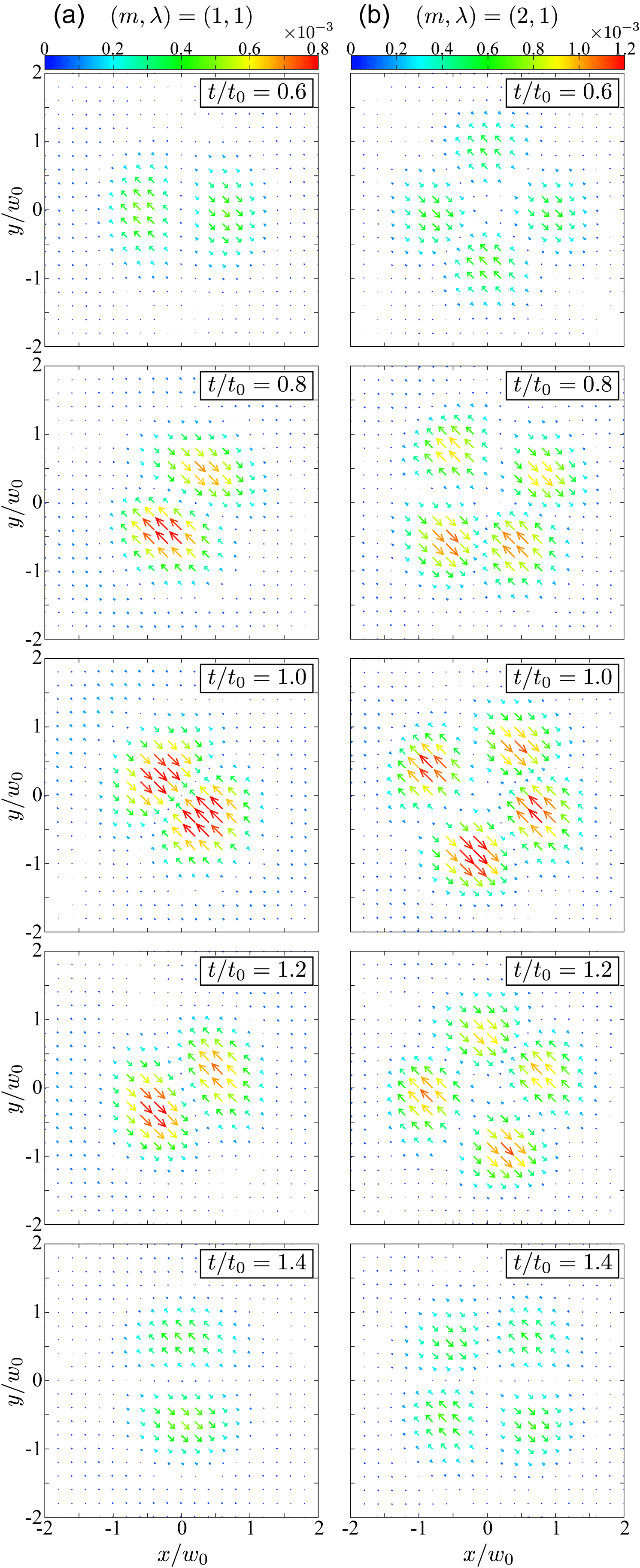}
    \caption{Snapshots of the local spin density in 2DEGs with $\alpha_{\rm R}k_{\rm F}=\beta_{\rm D}= 0.1~{\rm meV}$ after the irradiation of pulsed vortex beams with $(m,\lambda)=(1,1)$ (a) and $(2,1) (b)$.}
    \label{fig:timerd}
\end{figure}

Let us now examine the spin response in the case of $\alpha_{\rm R}=\beta_{\rm D}$, which significantly differs from that in the case of purely Rashba-type or Dresselhaus-type SOIs. Figure~\ref{fig:SU2-1} shows the linear response of the local spin to vortex beams with $(m,\lambda)=(1,1)$ and $(2,1)$. Similarly to the case of purely Rashba-type SOIs, the vortex beam causes spin polarization in the region where the intensity of the vortex beams peaks. However, as one sees from Fig.~\ref{fig:SU2-1}, the photo-induced spin moment predominantly points to one direction,  i.e., $[1\bar{1}0]$ direction, 
reflecting the symmetry at $\alpha_{\rm R}=\beta_{\rm D}$. In Fig.~\ref{fig:timerd}, we also plot the snapshots of the spin response to the pulsed vortex beam. At all times, the responded spins align to the $[1\bar{1}0]$ direction, and their intensity reflects a vortex shape influenced by the pulsed light. The movies demonstrating the real-time spin dynamics of 2DEGs with $\alpha_{\rm R}k_{\rm F}=\beta_{\rm D}k_{\rm F}=0.1~{\rm meV}$ are available in Supplemental Materilal~\cite{SM}. 

To understand the spin response at $\alpha_{\rm R}=\beta_{\rm D}$, let us consider the spin-current correlation function in Eq.~\eqref{eq:jscf} at $q\rightarrow 0$. By performing the spin rotation with the ${\rm SU}(2)$ matrix in Eqs.~\eqref{eq:HSU2}, the spin-current correlation function is recast into  
\begin{align}
\chi_{\mu\pm}(q)
\equiv &
\frac{\chi_{\mu x}(q)\pm \chi_{\mu y}(q)}{\sqrt{2}} \nn \\
=& \sum_k {\rm tr}_2
\left[ 
\tilde{G}(k)\hat{\tilde{S}}_{\mu}\tilde{G}(k+q)
\hat{\tilde{j}}_{\pm}({\bm k})
\right],
\end{align}
where $v_{\nu}\equiv \partial h_0({\bm k})/\partial k_{\nu}$, $\hat{\tilde{S}}_{\mu} \equiv U \hat{S}_{\mu}U^{\dag}$, and $\tilde{G}(k)\equiv UG(k)U^{\dag}= [i\varepsilon_n-\tilde{H}({\bm k})]^{-1}$. As shown in Eq.~\eqref{eq:HSU2}, the transformed Hamiltonian reduces to the one-dimensional SOI form, $k_+\tilde{\sigma_z}$, and the associated current operator is given by 
$\hat{\tilde{j}}_{\pm}({\bm k})
= -\frac{e \hbar k_+}{m_{\rm eff}} + (1\pm 1)e\alpha \tilde{\sigma}_z$. It is straightforward to show that on the transformed basis, the spin-current correlation function, ${\chi}_{\mu +}$, remains finite, while the other components ${\chi}_{\mu -}$ vanish. As a result, the spin response on the transformed spin basis reads $\tilde{S}_x = \tilde{S}_y=0$ and 
$\tilde{S}_{z}(q) = \tilde{\Upsilon}_{z+}(q)[E_x(q)+E_y(q)]/\sqrt{2}$, where $\tilde{\Upsilon}_{z+}(q)\equiv \tilde{\chi}_{z+}(q)/i\omega$. Let $R$ be the ${\rm SO}(3)$ matrix associated with $U$ in Eq.~\eqref{eq:SU2}. Using this matrix, the spin operators in the transformed spin basis are related to those in the original spin basis as $\tilde{S}_{\mu} = S_{\nu}R_{\nu\mu}$. Therefore, the spin response at $\alpha_{\rm R}=\beta_{\rm D}$ always satisfies 
\begin{align}
{\bm S}({\bm x},t) \approx S({\bm x},t)(1,-1,0).
\label{eq:spin-su2}
\end{align}
The amplitude is proportional to the sum of the electric field in the plane as
\begin{align}
S({\bm x},t) = \int dt^{\prime}\int d{\bm x}^{\prime}\tilde{\Upsilon}_{z+}({\bm x}-{\bm x}^{\prime},t-t^{\prime})E_+({\bm x}^{\prime},t^{\prime}),
\end{align}
where $E_{+}\equiv (E_x+E_y)/\sqrt{2}$. The numerical results shown in Figs.~\ref{fig:SU2-1} and \ref{fig:timerd} align with Eq.~\eqref{eq:spin-su2}, where all the spins are oriented to the $[1\bar{1}0]$ direction at all times.

\subsection{Orbital angular momentum dependence of spin response}
\begin{figure}[t!]
    \includegraphics[width=80mm]{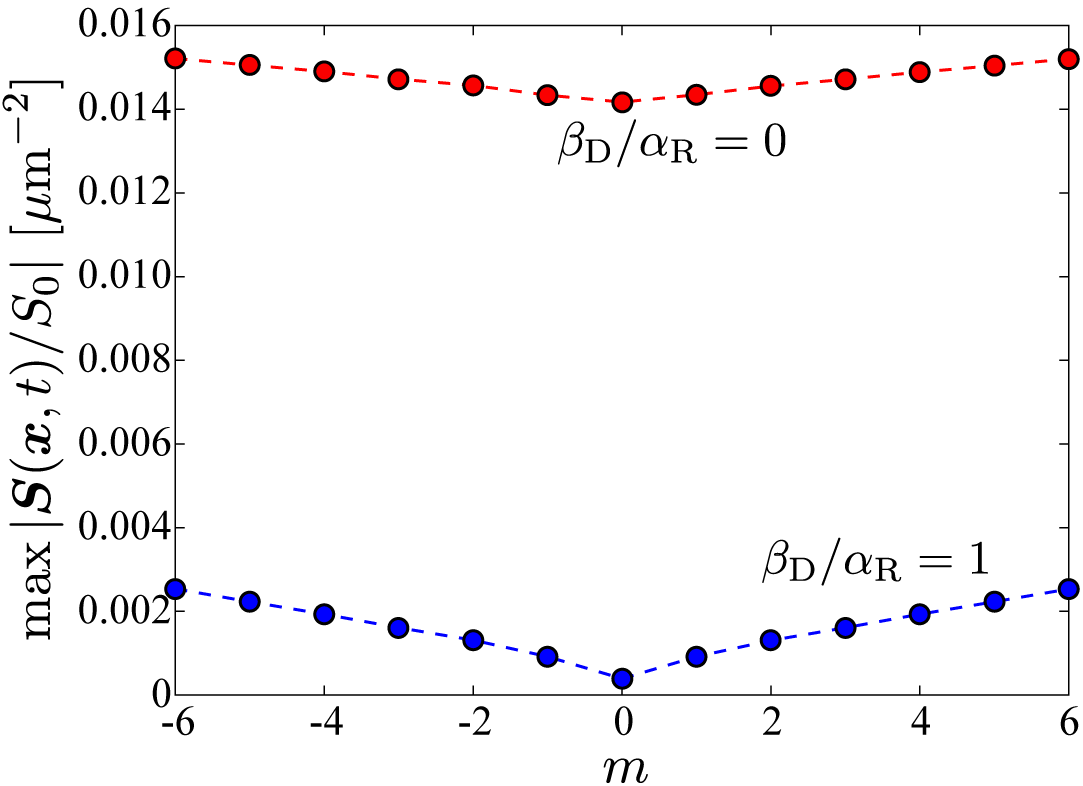}
    \caption{Maximal value of local spin polarization, $\max|S({\bm x},t)/S_0|$, as a function of the OAM of light, $m \in [-6,6]$. The red (blue) circles represent the results for $\beta_{\rm D}/\alpha_{\rm R}=0$ ($\beta_{\rm D}/\alpha_{\rm R}=1$). Here we set $\alpha_{\rm R}k_{\rm F}=0.1~{\rm meV}$ and $T/\varepsilon_{\rm F}=0.1$. \red{We note that the total spin is zero, $\int {\bm S}({\bm x},t)d{\bm x}={\bm 0}$, for $\alpha_{\rm R}=\beta_{\rm D}$.}}
    \label{fig:me}
\end{figure}

Lastly, we examine the OAM dependencies of the optical spin response in spin-orbit coupled electron gases. In Fig.~\ref{fig:me}, we plot the \red{numerically computed} maximal spin polarization, $\max|{\bm S}({\bm x},t)|/S_0$, as a function of the OAM of light, $m\in [-6,6]$. 
Here, $\max|{\bm S}({\bm x},t)|$ means the maximum value of \red{the laser-driven local spin magnitude} $|{\bm S}({\bm x},t)|$ over the entire range of $({\bm x}, t)$. 
In the case of the Rashba-type SOI ($\beta_{\rm D}/\alpha_{\rm R}=0$), the conventional Gaussiam beam, corresponding to optical vortices with $m=0$, causes nonvanishing spin polarization and the maximal spin polarization gradually increases with $|m|$. In contrast, the spins do not respond to the Gaussian beams when $\alpha_{\rm R}=\beta_{\rm D}$. This is consistent with the previous work~\cite{dey25}, where the spins of the 2DEG at the ${\rm SU}(2)$ symmetric point do not respond to the applied DC field, resulting in zero spin polarization.

It is also worth mentioning that for Gaussian beams, the maximal spin polarization at $\alpha_{\rm R}=\beta_{\rm D}$ is several orders of magnitude smaller compared to that in the case of $\beta_{\rm D}/\alpha_{\rm R}=0$. This is attributed to the ${\rm SU}(2)$ symmetry emerging at $\alpha_{\rm R}=\beta_{\rm D}$, where the effective magnetic field ${\bm b}_{\rm eff}({\bm k}) = \alpha_{\rm R}(k_x+k_y)(1,-1,0)$, defined in Eq.~\eqref{eq:hk}, causes unidirectional spin alignment. The intensity profile of conventional Gaussian beams peaks around $q=0$ in the momentum space, implying that the spin response is governed by the spatially uniform component of the spin-current response function, i.e., 
$S_{\mu}({\bm x},\omega)\approx \Upsilon_{\mu\nu}({\bm q}={\bm 0},\omega)E_{\nu}({\bm q},\omega)$. On the other hand, an optical vortex with nonzero $m$ is a structured light, and its intensity profile is a doughnut-like shape in the transverse plane. The spin response of 2DEGs peaks in the doughnut-shaped region, and the radius of the doughnut increases with $|m|$. For the vortex beam with a larger $|m|$, the contributions of the larger $|{\bm q}|$ to the spin-current correlation functions become significant. \red{As shown in Figs.~\ref{fig:SU2-1} and \ref{fig:timerd}, vortex beams with nonzero $m$ can induce a locally nontrivial spin texture while keeping the total spin at zero, i.e., 
\begin{align}
    \int {\bm S}({\bm x},t) d{\bm x}= 0,
    \label{eq:totalspin_zero}
\end{align}
under spin-helix conditions, $\alpha_{\rm R}=\beta_{\rm D}$. This results in} the strong OAM dependence of the local spin polarization. As shown in Fig.~\ref{fig:me}, the maximal spin polarization at $\alpha_{\rm R}=\beta_{\rm D}$ increases with increasing $|m|$ and reaches $0.002~\mu{\rm m}^{-2}$ for $E_0 = 0.1~{\rm kV}/{\rm cm}$. These results demonstrate that vortex beams enable ultrafast optical manipulation of spins even at the ${\rm SU}(2)$ symmetric point $\alpha_{\rm R}=\beta_{\rm D}$ for the realistic values of the AC electric field, $E_0$.

\section{Conclusion}

In this paper, we have examined the spin response of spin-orbit-coupled two-dimensional electron systems to optical vortices. We have performed numerical calculations on the spin response using linear response theory, considering both the case involving only the Rashba spin-orbit interaction and the case including both the Rashba and Dresselhaus spin-orbit interactions. We have demonstrated that optical vortices induce spin excitation through their spin-orbit coupling. In purely Rashba-type spin-orbit interaction, the excited spin textures are characterized by the topological charge of vortex beams, indicating that the orbital angular momentum of light can be transferred to the electron system. When Rashba and Dresselhaus spin-orbit interactions are of equal strength, the ${\rm SU}(2)$ symmetry prevents the imprinting of the spin textures characterized by the topological charge and results in unidirectional spin alignment at all times. 

We have also demonstrated that the magnitude of spin polarizations by vortex beams increases with the orbital angular momentum of light. In GaAs/AlGaAs heterostructures, THz vortex beams with an electric field of $E_0=0.1~{\rm kV}/{\rm cm}$ generates the spin polarization of about $0.015 \times (\hbar/2)$ per $1~\mu{\rm m}^2$ in purely Rashba-type spin-orbit interactions. In the equal strength of Rashba and Dresselhaus spin-orbit interactions, the conventional Gaussian beam, corresponding to the vortex beam with $m=0$, cannot induce the spin polarization. However, the spin polarization sharply depends on the orbital angular momentum of light, enabling spin polarizations by THz vortex beams with an electric field of $E_0\sim 1$-$10~{\rm kV}/{\rm cm}$. 

These results indicate that the pump beam with nonzero orbital angular momentum imprints the characteristic spin textures in 2DEGs. 
Our findings therefore demonstrate a high potential of optical vortex pulses for ultrafast spin manipulation in spin-orbit-coupled electrons.

\red{In this paper, the numerical calculations rely on linear response theory to aim at capturing the central aspect of spin manipulation using vortex beams. However, the retardation effect caused by the dipolar coupling of spins and a more precise treatment of impurity effects--beyond the weak impurity approximation--may play an important role when discussing spin dynamics in a realistic situation. These effects fall outside the scope of our current theory and remain as potential future work.}

Finally, we comment on a possible pump-probe experimental method of detecting the photo-induced spin textures.
The in-plane spin textures can be detected using the longitudinal or transverse magneto-optical Kerr effect~\cite{qiu99,qiu00}. A linearly polarized probe beam is incident at an angle to the normal direction of the 2DEGs, and the reflected light captures information about the in-plane magnetization through the longitudinal Kerr rotation. \red{Using the magneto-optical Kerr microscopy, the magnetization and magnetic domains in a material with in-plane anisotropy has been be obtained~\cite{zhou}. Another way is to induce the precession mode of in-plane electron spins by applying a static magnetic field. For example, it has recently been demonstrated that the higher-order spatial structure of photon polarization can be coherently transferred to an electron system in GaAs/AlGaAs quantum wells, imprinting spin textures varying along the azimuthal coordinate~\cite{mat24}. When a static magnetic field is applied in a plane, such as in the $x$-direction, individual spins start to precess around the $x$-axis. In Ref.~\cite{mat24}, the periodic time evolution of $S_z$ reflecting the spatial structure of photon polarization has been measured by time- and spatially-resolved magneto-optical Kerr rotation. Therefore, combining spatially-resolved Kerr rotation with a static magnetic field may reveal the characteristic spin textures induced by vortex beams.}

\begin{acknowledgments}
We thank Ken Morita for useful comments on the detection of in-plane spin textures. This work was supported by JST CREST (Grant No.~JPMJCR24R5), JSPS KAKENHI (Grants No.~JP23K20828, No.~JP23K22492, No.~JP25H00599, No.~JP25H00609, No.~JP25K07227, No.~JP25K07198, and No.~JP25K22011), a Grant-in-Aid for Transformative Research Areas (A) ``Correlation Design Science'' (Grant No.~JP25H01250 and No.~JP25H01251), ``Evolution of Chiral Materials Science using Helical Light Fields'' (Grant No. JP22H05131, JP23H04576 and JP25H01609), and ``Chimera Quasiparticles for Novel Condensed-Matter Science'' (Grant No. JP25H02112) from JSPS of Japan, \red{and JSPS Program for Forming J-PEAKS (Grant No.~JPJS00420230002)}.
\end{acknowledgments}

\appendix

\section{Impurity vertex correction}
\label{sec:vertex}

\begin{figure}[t!]
    \includegraphics[width=85mm]{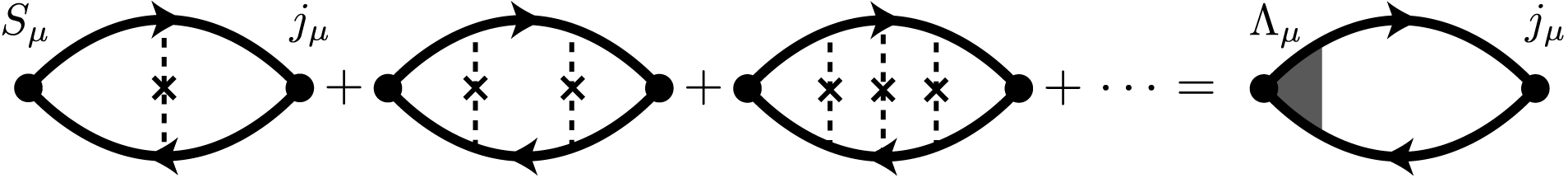}
    \caption{Diagrammatic representation of ladder-type impurity scattering for  vertex corrections ($\Lambda_{\mu}$) in spin-current response functions. The thick curves denote the Green's function $G(k)$, and the dashed lines connected to the symbol ``$\times$'' represent impurity scattering. 
    }
    \label{fig:diagram}
\end{figure}

In this Appendix, we discuss the effect of the impurity vertex corrections on the spin response of 2DEGs. The equation for the renormalized vertex functions that contains the product of retarded and advanced Green's functions takes the form
\begin{align}
\hat{\Lambda}_{\mu} = \hat{S}_{\mu} + n_{\rm imp}u^2_0\int \frac{d^2{\bm k}}{(2\pi)^2}
G^{\rm R}(k)\hat{\Lambda}_{\mu} G^{\rm A}(k+q).
\label{eq:vc}
\end{align}
These vertex functions incorporate the sum of the usual ladder diagrams as shown in Fig.~\ref{fig:diagram}.

We start by deriving the expression of the relaxation rate $\Gamma(\varepsilon)$. Let us define the density of states per spin as
\begin{align}
N(\varepsilon)= \frac{1}{2}\sum_s \int \frac{d^2k}{(2\pi)^2}
\delta(\varepsilon-E_s({\bm k})).
\end{align}
The relaxation rate is obtained by calculating Eq.~\eqref{eq:sigma} as
\begin{align}
\Gamma(\varepsilon) = \pi n_{\rm imp}u^2_0 N(\varepsilon).
\label{eq:gamma}
\end{align}

To solve Eq.~\eqref{eq:vc}, we take the following form of the vortex function: $\hat{\Lambda}_{\mu}(q) = \lambda^{\mu}_0(q) + \sum_{\nu} \lambda^{\mu}_{\nu}(q)\sigma_{\nu}$. Substituting this into Eq.~\eqref{eq:vc}, one obtains the coupled equations for $\lambda^{\mu}_0$ and $\lambda^{\mu}_{\nu}$? as 
\begin{gather}
A^+_{0}(q)\lambda_0^{\mu}(q)
+ A_{0\nu}(q)\lambda^{\mu}_{\nu}(q) = 0, 
\label{eq:E0} \\
-\frac{1}{2}\delta_{\mu \nu}
= A^-_{\nu 0}(q)\lambda^{\mu}_0(q)
+\left[ A_{0}(q)\delta_{\nu\eta}(q)+ A_{\nu\eta}(q)\right]\lambda^{\mu}_{\eta}(q),
\label{eq:Em}
\end{gather}
where the repeated Greek indices ($\nu,\eta,\tau$) imply the sum over $x,y,z$ and we introduce the shorthanded notations, $k\equiv ({\bm k},\varepsilon)$ and $q\equiv ({\bm q},\omega)$.
The coefficients are defined as 
\begin{align}
&A^{\pm}_0(q)= \Pi_{00}(q)\pm \sum_{\mu}\Pi_{\mu\mu}(q) -1, 
\label{eq:A00} \\
&A_{0\nu}(q)= \Pi_{0\nu}(q) + \Pi_{\nu 0}(q) 
+i\epsilon_{\nu\eta\tau} \Pi_{\eta\tau}(q), 
\label{eq:A0m} \\
&A_{\nu \eta}(q) = 
\Pi_{\nu\eta}(q) + \Pi_{\eta\nu}(q) + i\epsilon_{\nu\eta\tau}
\left[ \Pi_{0\tau}(q)
-\Pi_{\tau 0}(q)\right],
\label{eq:Amn}
\end{align}
and $A_{\nu 0} (q) = A_{0\nu}(q)$, where $i=0, x, y, z$ and $\epsilon_{ijk}$ is the Levi-Civita symbol. 
In Eqs.~\eqref{eq:A00}-\eqref{eq:Amn}, we have introduced the functions represented by
\beq
\Pi_{ij}(q) = n_{\rm imp}u^2_0 \int \frac{d^2{\bm k}}{(2\pi)^2} 
G^{\rm R}_i(k)G^{\rm A}_j(k+q),
\label{eq:Pi_ij}
\eeq
where $i,j=0,x,y,z$. 

Here we consider the vertex correction in the long wavelength limit, $q\ell \ll 1$, where $\ell = v_{\rm F}\tau$ is the mean-free-path of electrons and $v_{\rm F}$ is the Fermi velocity.  In the whole $q$ regime, the main contribution of $\Pi_{ij}(q)$ to the spin-current response function is located around the inverse of the beam waist, $q\sim 1/w_0$. THz vortex beams satisfy the condition $q\ell \ll 1$. Let us rewrite the retarded and advanced Green's functions, $G^{\rm R}(k)$ and $G^{\rm A}(k)$, as 
\beq
G^{{\rm R}/{\rm A}}(k) \equiv G_0^{{\rm R}/{\rm A}}(k) 
+ {G}^{{\rm R}/{\rm A}}_{\mu}(k)\sigma_{\mu}.
\label{eq:GRA}
\eeq
These are obtained from Eqs.~\eqref{eq:G} and \eqref{eq:Gs} with analytic continuation $i\varepsilon_n\rightarrow \varepsilon + i0_+$, where an infinitesimal constant $0_+$ is absorbed into the relaxation rate $\Gamma$. The scalar component is an even function in ${\bm k}$ as $G_0({\bm k},\varepsilon)= G_0(-{\bm k},\varepsilon)$, while the vectorial component is odd, $G_{\mu}({\bm k},\varepsilon) = -G_{\mu}(-{\bm k},\varepsilon)$. We also note that the three types of SOIs ($\beta_{\rm D}/\alpha_{\rm R}=0$, $\alpha_{\rm R}/\beta_{\rm D}=0$, and $\beta_{\rm D}/\alpha_{\rm R}=1$) hold $\Pi_{xx}=\Pi_{yy}$. Equations~\eqref{eq:E0} and \eqref{eq:Em} can be solved using these relations on the Green's functions and $\Pi_{ij}$. The solution of the expansion coefficients for $\Lambda_x$ in the limit of $q\ell \ll 1$ reads 
\begin{align}
\begin{pmatrix}
\lambda^{x}_{x} \\ \lambda^{x}_{y}
\end{pmatrix}
= \frac{\hbar}{2}\frac{1}{(1-\Pi_{00})^2-\Pi_{xy}\Pi_{yx}}
\begin{pmatrix}
1-\Pi_{00} \\ \Pi_{yx}
\end{pmatrix},
\end{align}
and $\lambda^{x}_{0}=\lambda^x_z=0$. 
The off-diagonal components are $\Pi_{xy}=\Pi_{yx}=0$ for the Rashba-type SOI and Dresslhaus-type SOI and $\Pi_{xy}=\Pi_{yx} \approx \Pi_{00}$ for $\Gamma\ll \varepsilon_{\rm F}$. The vertex function, $\Lambda_y$, is obtained in a similar way. 
As a result, the spin vertex functions reduce to the following form,
\begin{align}
\hat{\Lambda}_{\mu} = \sum_{\nu=x,y}\zeta_{\mu\nu} \hat{S}_{\nu}.
\end{align}
for $\mu=x,y$ and $\hat{\Lambda}_z=0$. 

The expressions for $\zeta_{\mu\nu}$ are obtained by evaluating $\Pi_{00}$ and $\Pi_{xy}$ at weak impurity concentrations. In purely Rashba-type SOIs, the latter function vanishes, i.e., $\Pi_{xy}=0$. The function at the Fermi energy, $\Pi_{00}(\omega)\equiv \Pi_{00}(\varepsilon=0,q=0,\omega)$, is obtained by substituting Eq.~\eqref{eq:GRA} into Eq.~\eqref{eq:Pi_ij} as 
\begin{align}
\Pi_{00}(\omega) =& \frac{n_{\rm imp}u^2_0}{4} 
\int \frac{d^2{\bm k}}{(2\pi)^2}
\sum_{s,s^{\prime}}G^{\rm R}_{s}({\bm k},0)G^{\rm A}_{s}({\bm k},\omega),
\label{eq:pi00}
\end{align}
For $\omega \tau \ll 1$, this function is recast into 
\begin{align}
\Pi_{00} = \frac{\pi n_{\rm imp}u^2_0}{2\Gamma}
\left[
N(0) + C\Gamma \int \frac{d^2{\bm k}}{(2\pi)^2} 
\prod_{s}\delta \left(\varepsilon-E_{s}(k)\right)
\right],
\end{align}
where $C=2\sqrt{2}\pi$.
The first term reduces to $1/2$ according to Eq.~\eqref{eq:gamma}. The second term yields nonzero contributions from the region where two split Fermi surfaces intersect, which is a minor correction on the order of $O(\Gamma/\varepsilon_{\rm F})$. Therefore, the function becomes
$\Pi_{00}=\frac{1}{2}$ and the renormalization factors are given by 
\begin{align}
\zeta _{\mu\nu} = 2\delta_{\mu\nu}.
\end{align}

In the main text, however, we consider the limit of weak impurity concentration, $\omega \tau \sim \Omega \tau \gg 1$. In this regime, the function, $\Pi_{00}$ can be evaluated by using 
$G^{\rm R}_s({\bm k},\varepsilon) = P\frac{1}{\varepsilon-E_s}-i\pi \delta(\varepsilon-E_s)$ for $\Gamma\rightarrow 0$, where $P$ denotes the Cauchy principal part. Substituting this into Eq.~\eqref{eq:pi00}, one finds ${\rm Re}\Pi_{00}=0$ and $-{\rm Im}\Pi_{00}\propto \frac{1}{\omega}$. This indicates that in the limit of $\omega \tau \gg 1$, the spin vertex function reduces to the bare spin operator, $\hat{\Lambda}_{\mu} \approx \hat{S}_{\mu}$. While we have considered Rashba-type SOIs, the same conclusion is valid for ${\rm SU}(2)$ symmetric SOIs as well.

\bibliography{reference}

\begin{thebibliography}{86}%
\makeatletter
\providecommand \@ifxundefined [1]{%
 \@ifx{#1\undefined}
}%
\providecommand \@ifnum [1]{%
 \ifnum #1\expandafter \@firstoftwo
 \else \expandafter \@secondoftwo
 \fi
}%
\providecommand \@ifx [1]{%
 \ifx #1\expandafter \@firstoftwo
 \else \expandafter \@secondoftwo
 \fi
}%
\providecommand \natexlab [1]{#1}%
\providecommand \enquote  [1]{``#1''}%
\providecommand \bibnamefont  [1]{#1}%
\providecommand \bibfnamefont [1]{#1}%
\providecommand \citenamefont [1]{#1}%
\providecommand \href@noop [0]{\@secondoftwo}%
\providecommand \href [0]{\begingroup \@sanitize@url \@href}%
\providecommand \@href[1]{\@@startlink{#1}\@@href}%
\providecommand \@@href[1]{\endgroup#1\@@endlink}%
\providecommand \@sanitize@url [0]{\catcode `\\12\catcode `\$12\catcode `\&12\catcode `\#12\catcode `\^12\catcode `\_12\catcode `\%12\relax}%
\providecommand \@@startlink[1]{}%
\providecommand \@@endlink[0]{}%
\providecommand \url  [0]{\begingroup\@sanitize@url \@url }%
\providecommand \@url [1]{\endgroup\@href {#1}{\urlprefix }}%
\providecommand \urlprefix  [0]{URL }%
\providecommand \Eprint [0]{\href }%
\providecommand \doibase [0]{https://doi.org/}%
\providecommand \selectlanguage [0]{\@gobble}%
\providecommand \bibinfo  [0]{\@secondoftwo}%
\providecommand \bibfield  [0]{\@secondoftwo}%
\providecommand \translation [1]{[#1]}%
\providecommand \BibitemOpen [0]{}%
\providecommand \bibitemStop [0]{}%
\providecommand \bibitemNoStop [0]{.\EOS\space}%
\providecommand \EOS [0]{\spacefactor3000\relax}%
\providecommand \BibitemShut  [1]{\csname bibitem#1\endcsname}%
\let\auto@bib@innerbib\@empty
\bibitem [{\citenamefont {Allen}\ \emph {et~al.}(1992)\citenamefont {Allen}, \citenamefont {Beijersbergen}, \citenamefont {Spreeuw},\ and\ \citenamefont {Woerdman}}]{Allen_1992}%
  \BibitemOpen
  \bibfield  {author} {\bibinfo {author} {\bibfnamefont {L.}~\bibnamefont {Allen}}, \bibinfo {author} {\bibfnamefont {M.~W.}\ \bibnamefont {Beijersbergen}}, \bibinfo {author} {\bibfnamefont {R.~J.~C.}\ \bibnamefont {Spreeuw}},\ and\ \bibinfo {author} {\bibfnamefont {J.~P.}\ \bibnamefont {Woerdman}},\ }\bibfield  {title} {\bibinfo {title} {{Orbital angular momentum of light and the transformation of Laguerre-Gaussian laser modes}},\ }\href {https://doi.org/10.1103/PhysRevA.45.8185} {\bibfield  {journal} {\bibinfo  {journal} {Phys. Rev. A}\ }\textbf {\bibinfo {volume} {45}},\ \bibinfo {pages} {8185} (\bibinfo {year} {1992})}\BibitemShut {NoStop}%
\bibitem [{\citenamefont {He}\ \emph {et~al.}(1995)\citenamefont {He}, \citenamefont {Friese}, \citenamefont {Heckenberg},\ and\ \citenamefont {Rubinsztein-Dunlop}}]{he95}%
  \BibitemOpen
  \bibfield  {author} {\bibinfo {author} {\bibfnamefont {H.}~\bibnamefont {He}}, \bibinfo {author} {\bibfnamefont {M.~E.~J.}\ \bibnamefont {Friese}}, \bibinfo {author} {\bibfnamefont {N.~R.}\ \bibnamefont {Heckenberg}},\ and\ \bibinfo {author} {\bibfnamefont {H.}~\bibnamefont {Rubinsztein-Dunlop}},\ }\bibfield  {title} {\bibinfo {title} {{Direct Observation of Transfer of Angular Momentum to Absorptive Particles from a Laser Beam with a Phase Singularity}},\ }\href {https://doi.org/10.1103/PhysRevLett.75.826} {\bibfield  {journal} {\bibinfo  {journal} {Phys. Rev. Lett.}\ }\textbf {\bibinfo {volume} {75}},\ \bibinfo {pages} {826} (\bibinfo {year} {1995})}\BibitemShut {NoStop}%
\bibitem [{\citenamefont {Kuga}\ \emph {et~al.}(1997{\natexlab{a}})\citenamefont {Kuga}, \citenamefont {Torii}, \citenamefont {Shiokawa}, \citenamefont {Hirano}, \citenamefont {Shimizu},\ and\ \citenamefont {Sasada}}]{kug97}%
  \BibitemOpen
  \bibfield  {author} {\bibinfo {author} {\bibfnamefont {T.}~\bibnamefont {Kuga}}, \bibinfo {author} {\bibfnamefont {Y.}~\bibnamefont {Torii}}, \bibinfo {author} {\bibfnamefont {N.}~\bibnamefont {Shiokawa}}, \bibinfo {author} {\bibfnamefont {T.}~\bibnamefont {Hirano}}, \bibinfo {author} {\bibfnamefont {Y.}~\bibnamefont {Shimizu}},\ and\ \bibinfo {author} {\bibfnamefont {H.}~\bibnamefont {Sasada}},\ }\bibfield  {title} {\bibinfo {title} {{Novel Optical Trap of Atoms with a Doughnut Beam}},\ }\href {https://doi.org/10.1103/PhysRevLett.78.4713} {\bibfield  {journal} {\bibinfo  {journal} {Phys. Rev. Lett.}\ }\textbf {\bibinfo {volume} {78}},\ \bibinfo {pages} {4713} (\bibinfo {year} {1997}{\natexlab{a}})}\BibitemShut {NoStop}%
\bibitem [{\citenamefont {O'Neil}\ \emph {et~al.}(2002{\natexlab{a}})\citenamefont {O'Neil}, \citenamefont {MacVicar}, \citenamefont {Allen},\ and\ \citenamefont {Padgett}}]{nei02}%
  \BibitemOpen
  \bibfield  {author} {\bibinfo {author} {\bibfnamefont {A.~T.}\ \bibnamefont {O'Neil}}, \bibinfo {author} {\bibfnamefont {I.}~\bibnamefont {MacVicar}}, \bibinfo {author} {\bibfnamefont {L.}~\bibnamefont {Allen}},\ and\ \bibinfo {author} {\bibfnamefont {M.~J.}\ \bibnamefont {Padgett}},\ }\bibfield  {title} {\bibinfo {title} {{Intrinsic and Extrinsic Nature of the Orbital Angular Momentum of a Light Beam}},\ }\href {https://doi.org/10.1103/PhysRevLett.88.053601} {\bibfield  {journal} {\bibinfo  {journal} {Phys. Rev. Lett.}\ }\textbf {\bibinfo {volume} {88}},\ \bibinfo {pages} {053601} (\bibinfo {year} {2002}{\natexlab{a}})}\BibitemShut {NoStop}%
\bibitem [{\citenamefont {MacDonald}\ \emph {et~al.}(2002)\citenamefont {MacDonald}, \citenamefont {Paterson}, \citenamefont {Volke-Sepulveda}, \citenamefont {Arlt}, \citenamefont {Sibbett},\ and\ \citenamefont {Dholakia}}]{mac02}%
  \BibitemOpen
  \bibfield  {author} {\bibinfo {author} {\bibfnamefont {M.~P.}\ \bibnamefont {MacDonald}}, \bibinfo {author} {\bibfnamefont {L.}~\bibnamefont {Paterson}}, \bibinfo {author} {\bibfnamefont {K.}~\bibnamefont {Volke-Sepulveda}}, \bibinfo {author} {\bibfnamefont {J.}~\bibnamefont {Arlt}}, \bibinfo {author} {\bibfnamefont {W.}~\bibnamefont {Sibbett}},\ and\ \bibinfo {author} {\bibfnamefont {K.}~\bibnamefont {Dholakia}},\ }\bibfield  {title} {\bibinfo {title} {{Creation and Manipulation of Three-Dimensional Optically Trapped Structures}},\ }\href {https://doi.org/10.1126/science.1069571} {\bibfield  {journal} {\bibinfo  {journal} {Science}\ }\textbf {\bibinfo {volume} {296}},\ \bibinfo {pages} {1101} (\bibinfo {year} {2002})}\BibitemShut {NoStop}%
\bibitem [{\citenamefont {Curtis}\ and\ \citenamefont {Grier}(2003)}]{cur03}%
  \BibitemOpen
  \bibfield  {author} {\bibinfo {author} {\bibfnamefont {J.~E.}\ \bibnamefont {Curtis}}\ and\ \bibinfo {author} {\bibfnamefont {D.~G.}\ \bibnamefont {Grier}},\ }\bibfield  {title} {\bibinfo {title} {{Structure of Optical Vortices}},\ }\href {https://doi.org/10.1103/PhysRevLett.90.133901} {\bibfield  {journal} {\bibinfo  {journal} {Phys. Rev. Lett.}\ }\textbf {\bibinfo {volume} {90}},\ \bibinfo {pages} {133901} (\bibinfo {year} {2003})}\BibitemShut {NoStop}%
\bibitem [{\citenamefont {Dholakia}\ and\ \citenamefont {\v{C}i\v{z}m\'{a}r}(2011)}]{dho11}%
  \BibitemOpen
  \bibfield  {author} {\bibinfo {author} {\bibfnamefont {K.}~\bibnamefont {Dholakia}}\ and\ \bibinfo {author} {\bibfnamefont {T.}~\bibnamefont {\v{C}i\v{z}m\'{a}r}},\ }\bibfield  {title} {\bibinfo {title} {{Shaping the future of manipulation}},\ }\href {https://doi.org/10.1038/nphoton.2011.80} {\bibfield  {journal} {\bibinfo  {journal} {Nat. Photonics}\ }\textbf {\bibinfo {volume} {5}},\ \bibinfo {pages} {335} (\bibinfo {year} {2011})}\BibitemShut {NoStop}%
\bibitem [{\citenamefont {Kuga}\ \emph {et~al.}(1997{\natexlab{b}})\citenamefont {Kuga}, \citenamefont {Torii}, \citenamefont {Shiokawa}, \citenamefont {Hirano}, \citenamefont {Shimizu},\ and\ \citenamefont {Sasada}}]{Kuga_1997}%
  \BibitemOpen
  \bibfield  {author} {\bibinfo {author} {\bibfnamefont {T.}~\bibnamefont {Kuga}}, \bibinfo {author} {\bibfnamefont {Y.}~\bibnamefont {Torii}}, \bibinfo {author} {\bibfnamefont {N.}~\bibnamefont {Shiokawa}}, \bibinfo {author} {\bibfnamefont {T.}~\bibnamefont {Hirano}}, \bibinfo {author} {\bibfnamefont {Y.}~\bibnamefont {Shimizu}},\ and\ \bibinfo {author} {\bibfnamefont {H.}~\bibnamefont {Sasada}},\ }\bibfield  {title} {\bibinfo {title} {{Novel Optical Trap of Atoms with a Doughnut Beam}},\ }\href {https://doi.org/10.1103/PhysRevLett.78.4713} {\bibfield  {journal} {\bibinfo  {journal} {Phys. Rev. Lett.}\ }\textbf {\bibinfo {volume} {78}},\ \bibinfo {pages} {4713} (\bibinfo {year} {1997}{\natexlab{b}})}\BibitemShut {NoStop}%
\bibitem [{\citenamefont {O'Neil}\ \emph {et~al.}(2002{\natexlab{b}})\citenamefont {O'Neil}, \citenamefont {MacVicar}, \citenamefont {Allen},\ and\ \citenamefont {Padgett}}]{Neil_2002}%
  \BibitemOpen
  \bibfield  {author} {\bibinfo {author} {\bibfnamefont {A.~T.}\ \bibnamefont {O'Neil}}, \bibinfo {author} {\bibfnamefont {I.}~\bibnamefont {MacVicar}}, \bibinfo {author} {\bibfnamefont {L.}~\bibnamefont {Allen}},\ and\ \bibinfo {author} {\bibfnamefont {M.~J.}\ \bibnamefont {Padgett}},\ }\bibfield  {title} {\bibinfo {title} {{Intrinsic and Extrinsic Nature of the Orbital Angular Momentum of a Light Beam}},\ }\href {https://doi.org/10.1103/PhysRevLett.88.053601} {\bibfield  {journal} {\bibinfo  {journal} {Phys. Rev. Lett.}\ }\textbf {\bibinfo {volume} {88}},\ \bibinfo {pages} {053601} (\bibinfo {year} {2002}{\natexlab{b}})}\BibitemShut {NoStop}%
\bibitem [{\citenamefont {Omatsu}\ \emph {et~al.}(2010)\citenamefont {Omatsu}, \citenamefont {Chujo}, \citenamefont {Miyamoto}, \citenamefont {Okida}, \citenamefont {Nakamura}, \citenamefont {Aoki},\ and\ \citenamefont {Morita}}]{oma10}%
  \BibitemOpen
  \bibfield  {author} {\bibinfo {author} {\bibfnamefont {T.}~\bibnamefont {Omatsu}}, \bibinfo {author} {\bibfnamefont {K.}~\bibnamefont {Chujo}}, \bibinfo {author} {\bibfnamefont {K.}~\bibnamefont {Miyamoto}}, \bibinfo {author} {\bibfnamefont {M.}~\bibnamefont {Okida}}, \bibinfo {author} {\bibfnamefont {K.}~\bibnamefont {Nakamura}}, \bibinfo {author} {\bibfnamefont {N.}~\bibnamefont {Aoki}},\ and\ \bibinfo {author} {\bibfnamefont {R.}~\bibnamefont {Morita}},\ }\bibfield  {title} {\bibinfo {title} {Metal microneedle fabrication using twisted light with spin},\ }\href {https://doi.org/10.1364/OE.18.017967} {\bibfield  {journal} {\bibinfo  {journal} {Opt. Express}\ }\textbf {\bibinfo {volume} {18}},\ \bibinfo {pages} {17967} (\bibinfo {year} {2010})}\BibitemShut {NoStop}%
\bibitem [{\citenamefont {Toyoda}\ \emph {et~al.}(2012)\citenamefont {Toyoda}, \citenamefont {Miyamoto}, \citenamefont {Aoki}, \citenamefont {Morita},\ and\ \citenamefont {Omatsu}}]{toy12}%
  \BibitemOpen
  \bibfield  {author} {\bibinfo {author} {\bibfnamefont {K.}~\bibnamefont {Toyoda}}, \bibinfo {author} {\bibfnamefont {K.}~\bibnamefont {Miyamoto}}, \bibinfo {author} {\bibfnamefont {N.}~\bibnamefont {Aoki}}, \bibinfo {author} {\bibfnamefont {R.}~\bibnamefont {Morita}},\ and\ \bibinfo {author} {\bibfnamefont {T.}~\bibnamefont {Omatsu}},\ }\bibfield  {title} {\bibinfo {title} {{Using Optical Vortex To Control the Chirality of Twisted Metal Nanostructures}},\ }\href {https://doi.org/10.1021/nl301347j} {\bibfield  {journal} {\bibinfo  {journal} {Nano Lett.}\ }\textbf {\bibinfo {volume} {12}},\ \bibinfo {pages} {3645} (\bibinfo {year} {2012})}\BibitemShut {NoStop}%
\bibitem [{\citenamefont {Toyoda}\ \emph {et~al.}(2013)\citenamefont {Toyoda}, \citenamefont {Takahashi}, \citenamefont {Takizawa}, \citenamefont {Tokizane}, \citenamefont {Miyamoto}, \citenamefont {Morita},\ and\ \citenamefont {Omatsu}}]{toy13}%
  \BibitemOpen
  \bibfield  {author} {\bibinfo {author} {\bibfnamefont {K.}~\bibnamefont {Toyoda}}, \bibinfo {author} {\bibfnamefont {F.}~\bibnamefont {Takahashi}}, \bibinfo {author} {\bibfnamefont {S.}~\bibnamefont {Takizawa}}, \bibinfo {author} {\bibfnamefont {Y.}~\bibnamefont {Tokizane}}, \bibinfo {author} {\bibfnamefont {K.}~\bibnamefont {Miyamoto}}, \bibinfo {author} {\bibfnamefont {R.}~\bibnamefont {Morita}},\ and\ \bibinfo {author} {\bibfnamefont {T.}~\bibnamefont {Omatsu}},\ }\bibfield  {title} {\bibinfo {title} {{Transfer of Light Helicity to Nanostructures}},\ }\href {https://doi.org/10.1103/PhysRevLett.110.143603} {\bibfield  {journal} {\bibinfo  {journal} {Phys. Rev. Lett.}\ }\textbf {\bibinfo {volume} {110}},\ \bibinfo {pages} {143603} (\bibinfo {year} {2013})}\BibitemShut {NoStop}%
\bibitem [{\citenamefont {Barreiro}\ \emph {et~al.}(2008)\citenamefont {Barreiro}, \citenamefont {Wei},\ and\ \citenamefont {Kwiat}}]{Barrerio_2008}%
  \BibitemOpen
  \bibfield  {author} {\bibinfo {author} {\bibfnamefont {J.~T.}\ \bibnamefont {Barreiro}}, \bibinfo {author} {\bibfnamefont {T.-C.}\ \bibnamefont {Wei}},\ and\ \bibinfo {author} {\bibfnamefont {P.~G.}\ \bibnamefont {Kwiat}},\ }\bibfield  {title} {\bibinfo {title} {{Beating the channel capacity limit for linear photonic superdense coding}},\ }\href {https://doi.org/10.1038/nphys919} {\bibfield  {journal} {\bibinfo  {journal} {Nat. Phys.}\ }\textbf {\bibinfo {volume} {4}},\ \bibinfo {pages} {282} (\bibinfo {year} {2008})}\BibitemShut {NoStop}%
\bibitem [{\citenamefont {Wang}\ \emph {et~al.}(2012)\citenamefont {Wang}, \citenamefont {Yang}, \citenamefont {Fazal}, \citenamefont {Ahmed}, \citenamefont {Yan}, \citenamefont {Huang}, \citenamefont {Ren}, \citenamefont {Yue}, \citenamefont {Dolinar}, \citenamefont {Tur} \emph {et~al.}}]{Wang_2012}%
  \BibitemOpen
  \bibfield  {author} {\bibinfo {author} {\bibfnamefont {J.}~\bibnamefont {Wang}}, \bibinfo {author} {\bibfnamefont {J.-Y.}\ \bibnamefont {Yang}}, \bibinfo {author} {\bibfnamefont {I.~M.}\ \bibnamefont {Fazal}}, \bibinfo {author} {\bibfnamefont {N.}~\bibnamefont {Ahmed}}, \bibinfo {author} {\bibfnamefont {Y.}~\bibnamefont {Yan}}, \bibinfo {author} {\bibfnamefont {H.}~\bibnamefont {Huang}}, \bibinfo {author} {\bibfnamefont {Y.}~\bibnamefont {Ren}}, \bibinfo {author} {\bibfnamefont {Y.}~\bibnamefont {Yue}}, \bibinfo {author} {\bibfnamefont {S.}~\bibnamefont {Dolinar}}, \bibinfo {author} {\bibfnamefont {M.}~\bibnamefont {Tur}}, \emph {et~al.},\ }\bibfield  {title} {\bibinfo {title} {{Terabit free-space data transmission employing orbital angular momentum multiplexing}},\ }\href {https://doi.org/10.1038/nphoton.2012.138} {\bibfield  {journal} {\bibinfo  {journal} {Nature photonics}\ }\textbf {\bibinfo {volume} {6}},\ \bibinfo {pages} {488} (\bibinfo {year} {2012})}\BibitemShut {NoStop}%
\bibitem [{\citenamefont {Bozinovic}\ \emph {et~al.}(2013)\citenamefont {Bozinovic}, \citenamefont {Yue}, \citenamefont {Ren}, \citenamefont {Tur}, \citenamefont {Kristensen}, \citenamefont {Huang}, \citenamefont {Willner},\ and\ \citenamefont {Ramachandran}}]{Bozinovic_2013}%
  \BibitemOpen
  \bibfield  {author} {\bibinfo {author} {\bibfnamefont {N.}~\bibnamefont {Bozinovic}}, \bibinfo {author} {\bibfnamefont {Y.}~\bibnamefont {Yue}}, \bibinfo {author} {\bibfnamefont {Y.}~\bibnamefont {Ren}}, \bibinfo {author} {\bibfnamefont {M.}~\bibnamefont {Tur}}, \bibinfo {author} {\bibfnamefont {P.}~\bibnamefont {Kristensen}}, \bibinfo {author} {\bibfnamefont {H.}~\bibnamefont {Huang}}, \bibinfo {author} {\bibfnamefont {A.~E.}\ \bibnamefont {Willner}},\ and\ \bibinfo {author} {\bibfnamefont {S.}~\bibnamefont {Ramachandran}},\ }\bibfield  {title} {\bibinfo {title} {{Terabit-scale orbital angular momentum mode division multiplexing in fibers}},\ }\href {https://doi.org/10.1126/science.1237861} {\bibfield  {journal} {\bibinfo  {journal} {Science}\ }\textbf {\bibinfo {volume} {340}},\ \bibinfo {pages} {1545} (\bibinfo {year} {2013})}\BibitemShut {NoStop}%
\bibitem [{\citenamefont {Quinteiro~Rosen}\ \emph {et~al.}(2022)\citenamefont {Quinteiro~Rosen}, \citenamefont {Tamborenea},\ and\ \citenamefont {Kuhn}}]{qui22}%
  \BibitemOpen
  \bibfield  {author} {\bibinfo {author} {\bibfnamefont {G.~F.}\ \bibnamefont {Quinteiro~Rosen}}, \bibinfo {author} {\bibfnamefont {P.~I.}\ \bibnamefont {Tamborenea}},\ and\ \bibinfo {author} {\bibfnamefont {T.}~\bibnamefont {Kuhn}},\ }\bibfield  {title} {\bibinfo {title} {{Interplay between optical vortices and condensed matter}},\ }\href {https://doi.org/10.1103/RevModPhys.94.035003} {\bibfield  {journal} {\bibinfo  {journal} {Rev. Mod. Phys.}\ }\textbf {\bibinfo {volume} {94}},\ \bibinfo {pages} {035003} (\bibinfo {year} {2022})}\BibitemShut {NoStop}%
\bibitem [{\citenamefont {Shintani}\ \emph {et~al.}(2016)\citenamefont {Shintani}, \citenamefont {Taguchi}, \citenamefont {Tanaka},\ and\ \citenamefont {Kawaguchi}}]{shintani_2016}%
  \BibitemOpen
  \bibfield  {author} {\bibinfo {author} {\bibfnamefont {K.}~\bibnamefont {Shintani}}, \bibinfo {author} {\bibfnamefont {K.}~\bibnamefont {Taguchi}}, \bibinfo {author} {\bibfnamefont {Y.}~\bibnamefont {Tanaka}},\ and\ \bibinfo {author} {\bibfnamefont {Y.}~\bibnamefont {Kawaguchi}},\ }\bibfield  {title} {\bibinfo {title} {{Spin and charge transport induced by a twisted light beam on the surface of a topological insulator}},\ }\href {https://doi.org/10.1103/PhysRevB.93.195415} {\bibfield  {journal} {\bibinfo  {journal} {Phys. Rev. B}\ }\textbf {\bibinfo {volume} {93}},\ \bibinfo {pages} {195415} (\bibinfo {year} {2016})}\BibitemShut {NoStop}%
\bibitem [{\citenamefont {Fujita}\ and\ \citenamefont {Sato}(2018)}]{fuj18}%
  \BibitemOpen
  \bibfield  {author} {\bibinfo {author} {\bibfnamefont {H.}~\bibnamefont {Fujita}}\ and\ \bibinfo {author} {\bibfnamefont {M.}~\bibnamefont {Sato}},\ }\bibfield  {title} {\bibinfo {title} {{Nonequilibrium Magnetic Oscillation with Cylindrical Vector Beams}},\ }\href {https://doi.org/10.1038/s41598-018-33651-0} {\bibfield  {journal} {\bibinfo  {journal} {Sci. Rep.}\ }\textbf {\bibinfo {volume} {8}},\ \bibinfo {pages} {15738} (\bibinfo {year} {2018})}\BibitemShut {NoStop}%
\bibitem [{\citenamefont {Fujita}\ \emph {et~al.}(2019)\citenamefont {Fujita}, \citenamefont {Tada},\ and\ \citenamefont {Sato}}]{fuj19}%
  \BibitemOpen
  \bibfield  {author} {\bibinfo {author} {\bibfnamefont {H.}~\bibnamefont {Fujita}}, \bibinfo {author} {\bibfnamefont {Y.}~\bibnamefont {Tada}},\ and\ \bibinfo {author} {\bibfnamefont {M.}~\bibnamefont {Sato}},\ }\bibfield  {title} {\bibinfo {title} {{Accessing electromagnetic properties of matter with cylindrical vector beams}},\ }\href {https://doi.org/10.1088/1367-2630/ab26d1} {\bibfield  {journal} {\bibinfo  {journal} {New J. Phys.}\ }\textbf {\bibinfo {volume} {21}},\ \bibinfo {pages} {073010} (\bibinfo {year} {2019})}\BibitemShut {NoStop}%
\bibitem [{\citenamefont {Takahashi}\ \emph {et~al.}(2018)\citenamefont {Takahashi}, \citenamefont {Proskurin},\ and\ \citenamefont {Kishine}}]{tak18}%
  \BibitemOpen
  \bibfield  {author} {\bibinfo {author} {\bibfnamefont {H.~T.}\ \bibnamefont {Takahashi}}, \bibinfo {author} {\bibfnamefont {I.}~\bibnamefont {Proskurin}},\ and\ \bibinfo {author} {\bibfnamefont {J.-i.}\ \bibnamefont {Kishine}},\ }\bibfield  {title} {\bibinfo {title} {{Landau Level Spectroscopy by Optical Vortex Beam}},\ }\href {https://doi.org/10.7566/JPSJ.87.113703} {\bibfield  {journal} {\bibinfo  {journal} {J. Phys. Soc. Jpn.}\ }\textbf {\bibinfo {volume} {87}},\ \bibinfo {pages} {113703} (\bibinfo {year} {2018})}\BibitemShut {NoStop}%
\bibitem [{\citenamefont {Mizushima}\ and\ \citenamefont {Sato}(2023)}]{mizushima_2023}%
  \BibitemOpen
  \bibfield  {author} {\bibinfo {author} {\bibfnamefont {T.}~\bibnamefont {Mizushima}}\ and\ \bibinfo {author} {\bibfnamefont {M.}~\bibnamefont {Sato}},\ }\bibfield  {title} {\bibinfo {title} {{Imprinting spiral Higgs waves onto superconductors with vortex beams}},\ }\href {https://doi.org/10.1103/PhysRevResearch.5.L042004} {\bibfield  {journal} {\bibinfo  {journal} {Phys. Rev. Res.}\ }\textbf {\bibinfo {volume} {5}},\ \bibinfo {pages} {L042004} (\bibinfo {year} {2023})}\BibitemShut {NoStop}%
\bibitem [{\citenamefont {Toda}\ \emph {et~al.}(2023)\citenamefont {Toda}, \citenamefont {Tsuchiya}, \citenamefont {Yamane}, \citenamefont {Morita}, \citenamefont {Oda}, \citenamefont {Kurosawa}, \citenamefont {Mertelj},\ and\ \citenamefont {Mihailovic}}]{tod23}%
  \BibitemOpen
  \bibfield  {author} {\bibinfo {author} {\bibfnamefont {Y.}~\bibnamefont {Toda}}, \bibinfo {author} {\bibfnamefont {S.}~\bibnamefont {Tsuchiya}}, \bibinfo {author} {\bibfnamefont {K.}~\bibnamefont {Yamane}}, \bibinfo {author} {\bibfnamefont {R.}~\bibnamefont {Morita}}, \bibinfo {author} {\bibfnamefont {M.}~\bibnamefont {Oda}}, \bibinfo {author} {\bibfnamefont {T.}~\bibnamefont {Kurosawa}}, \bibinfo {author} {\bibfnamefont {T.}~\bibnamefont {Mertelj}},\ and\ \bibinfo {author} {\bibfnamefont {D.}~\bibnamefont {Mihailovic}},\ }\bibfield  {title} {\bibinfo {title} {{Optical vortex induced spatio-temporally modulated superconductivity in a high-Tc cuprate}},\ }\href {https://doi.org/10.1364/OE.487041} {\bibfield  {journal} {\bibinfo  {journal} {Opt. Express}\ }\textbf {\bibinfo {volume} {31}},\ \bibinfo {pages} {17537} (\bibinfo {year} {2023})}\BibitemShut {NoStop}%
\bibitem [{\citenamefont {Yerzhakov}\ \emph {et~al.}(2024)\citenamefont {Yerzhakov}, \citenamefont {Yeh},\ and\ \citenamefont {Balatsky}}]{yer24}%
  \BibitemOpen
  \bibfield  {author} {\bibinfo {author} {\bibfnamefont {H.}~\bibnamefont {Yerzhakov}}, \bibinfo {author} {\bibfnamefont {T.-T.}\ \bibnamefont {Yeh}},\ and\ \bibinfo {author} {\bibfnamefont {A.}~\bibnamefont {Balatsky}},\ }\bibfield  {title} {\bibinfo {title} {{Induction of orbital currents and Kapitza stabilization in superconducting circuits with Laguerre-Gaussian microwave beams}},\ }\href {https://doi.org/10.1103/PhysRevB.110.144519} {\bibfield  {journal} {\bibinfo  {journal} {Phys. Rev. B}\ }\textbf {\bibinfo {volume} {110}},\ \bibinfo {pages} {144519} (\bibinfo {year} {2024})}\BibitemShut {NoStop}%
\bibitem [{\citenamefont {Yeh}\ \emph {et~al.}(2025{\natexlab{a}})\citenamefont {Yeh}, \citenamefont {Yerzhakov}, \citenamefont {Horn}, \citenamefont {Raghu},\ and\ \citenamefont {Balatsky}}]{yeh24}%
  \BibitemOpen
  \bibfield  {author} {\bibinfo {author} {\bibfnamefont {T.-T.}\ \bibnamefont {Yeh}}, \bibinfo {author} {\bibfnamefont {H.}~\bibnamefont {Yerzhakov}}, \bibinfo {author} {\bibfnamefont {L.~B.-V.}\ \bibnamefont {Horn}}, \bibinfo {author} {\bibfnamefont {S.}~\bibnamefont {Raghu}},\ and\ \bibinfo {author} {\bibfnamefont {A.}~\bibnamefont {Balatsky}},\ }\href {https://arxiv.org/abs/2407.15834} {\bibinfo {title} {{Structured light and induced vorticity in superconductors I: Linearly polarized light}}} (\bibinfo {year} {2025}{\natexlab{a}}),\ \Eprint {https://arxiv.org/abs/2407.15834} {arXiv:2407.15834 [cond-mat.supr-con]} \BibitemShut {NoStop}%
\bibitem [{\citenamefont {Yeh}\ \emph {et~al.}(2025{\natexlab{b}})\citenamefont {Yeh}, \citenamefont {Yerzhakov}, \citenamefont {Horn}, \citenamefont {Raghu},\ and\ \citenamefont {Balatsky}}]{yeh25}%
  \BibitemOpen
  \bibfield  {author} {\bibinfo {author} {\bibfnamefont {T.-T.}\ \bibnamefont {Yeh}}, \bibinfo {author} {\bibfnamefont {H.}~\bibnamefont {Yerzhakov}}, \bibinfo {author} {\bibfnamefont {L.~B.-V.}\ \bibnamefont {Horn}}, \bibinfo {author} {\bibfnamefont {S.}~\bibnamefont {Raghu}},\ and\ \bibinfo {author} {\bibfnamefont {A.}~\bibnamefont {Balatsky}},\ }\href {https://arxiv.org/abs/2412.00935} {\bibinfo {title} {{Structured light and induced vorticity in superconductors II: Quantum Print with Laguerre-Gaussian beam}}} (\bibinfo {year} {2025}{\natexlab{b}}),\ \Eprint {https://arxiv.org/abs/2412.00935} {arXiv:2412.00935 [cond-mat.supr-con]} \BibitemShut {NoStop}%
\bibitem [{\citenamefont {Kang}\ \emph {et~al.}(2025)\citenamefont {Kang}, \citenamefont {Kitamura},\ and\ \citenamefont {Morimoto}}]{kang25}%
  \BibitemOpen
  \bibfield  {author} {\bibinfo {author} {\bibfnamefont {D.}~\bibnamefont {Kang}}, \bibinfo {author} {\bibfnamefont {S.}~\bibnamefont {Kitamura}},\ and\ \bibinfo {author} {\bibfnamefont {T.}~\bibnamefont {Morimoto}},\ }\href {https://arxiv.org/abs/2504.11883} {\bibinfo {title} {{Quantum Optical Spanner: Twisting Superconductors with Vortex Beam via Higgs Mode}}} (\bibinfo {year} {2025}),\ \Eprint {https://arxiv.org/abs/2504.11883} {arXiv:2504.11883 [cond-mat.str-el]} \BibitemShut {NoStop}%
\bibitem [{\citenamefont {Fujita}\ and\ \citenamefont {Sato}(2017{\natexlab{a}})}]{sato1_2017}%
  \BibitemOpen
  \bibfield  {author} {\bibinfo {author} {\bibfnamefont {H.}~\bibnamefont {Fujita}}\ and\ \bibinfo {author} {\bibfnamefont {M.}~\bibnamefont {Sato}},\ }\bibfield  {title} {\bibinfo {title} {{Ultrafast generation of skyrmionic defects with vortex beams: Printing laser profiles on magnets}},\ }\href {https://doi.org/10.1103/PhysRevB.95.054421} {\bibfield  {journal} {\bibinfo  {journal} {Phys. Rev. B}\ }\textbf {\bibinfo {volume} {95}},\ \bibinfo {pages} {054421} (\bibinfo {year} {2017}{\natexlab{a}})}\BibitemShut {NoStop}%
\bibitem [{\citenamefont {Fujita}\ and\ \citenamefont {Sato}(2017{\natexlab{b}})}]{sato2_2017}%
  \BibitemOpen
  \bibfield  {author} {\bibinfo {author} {\bibfnamefont {H.}~\bibnamefont {Fujita}}\ and\ \bibinfo {author} {\bibfnamefont {M.}~\bibnamefont {Sato}},\ }\bibfield  {title} {\bibinfo {title} {{Encoding orbital angular momentum of light in magnets}},\ }\href {https://doi.org/10.1103/PhysRevB.96.060407} {\bibfield  {journal} {\bibinfo  {journal} {Phys. Rev. B}\ }\textbf {\bibinfo {volume} {96}},\ \bibinfo {pages} {060407} (\bibinfo {year} {2017}{\natexlab{b}})}\BibitemShut {NoStop}%
\bibitem [{\citenamefont {Sirenko}\ \emph {et~al.}(2019)\citenamefont {Sirenko}, \citenamefont {Marsik}, \citenamefont {Bernhard}, \citenamefont {Stanislavchuk}, \citenamefont {Kiryukhin},\ and\ \citenamefont {Cheong}}]{sir19}%
  \BibitemOpen
  \bibfield  {author} {\bibinfo {author} {\bibfnamefont {A.~A.}\ \bibnamefont {Sirenko}}, \bibinfo {author} {\bibfnamefont {P.}~\bibnamefont {Marsik}}, \bibinfo {author} {\bibfnamefont {C.}~\bibnamefont {Bernhard}}, \bibinfo {author} {\bibfnamefont {T.~N.}\ \bibnamefont {Stanislavchuk}}, \bibinfo {author} {\bibfnamefont {V.}~\bibnamefont {Kiryukhin}},\ and\ \bibinfo {author} {\bibfnamefont {S.-W.}\ \bibnamefont {Cheong}},\ }\bibfield  {title} {\bibinfo {title} {{Terahertz Vortex Beam as a Spectroscopic Probe of Magnetic Excitations}},\ }\href {https://doi.org/10.1103/PhysRevLett.122.237401} {\bibfield  {journal} {\bibinfo  {journal} {Phys. Rev. Lett.}\ }\textbf {\bibinfo {volume} {122}},\ \bibinfo {pages} {237401} (\bibinfo {year} {2019})}\BibitemShut {NoStop}%
\bibitem [{\citenamefont {Sirenko}\ \emph {et~al.}(2021)\citenamefont {Sirenko}, \citenamefont {Marsik}, \citenamefont {Bugnon}, \citenamefont {Soulier}, \citenamefont {Bernhard}, \citenamefont {Stanislavchuk}, \citenamefont {Xu},\ and\ \citenamefont {Cheong}}]{sir21}%
  \BibitemOpen
  \bibfield  {author} {\bibinfo {author} {\bibfnamefont {A.~A.}\ \bibnamefont {Sirenko}}, \bibinfo {author} {\bibfnamefont {P.}~\bibnamefont {Marsik}}, \bibinfo {author} {\bibfnamefont {L.}~\bibnamefont {Bugnon}}, \bibinfo {author} {\bibfnamefont {M.}~\bibnamefont {Soulier}}, \bibinfo {author} {\bibfnamefont {C.}~\bibnamefont {Bernhard}}, \bibinfo {author} {\bibfnamefont {T.~N.}\ \bibnamefont {Stanislavchuk}}, \bibinfo {author} {\bibfnamefont {X.}~\bibnamefont {Xu}},\ and\ \bibinfo {author} {\bibfnamefont {S.-W.}\ \bibnamefont {Cheong}},\ }\bibfield  {title} {\bibinfo {title} {{Total Angular Momentum Dichroism of the Terahertz Vortex Beams at the Antiferromagnetic Resonances}},\ }\href {https://doi.org/10.1103/PhysRevLett.126.157401} {\bibfield  {journal} {\bibinfo  {journal} {Phys. Rev. Lett.}\ }\textbf {\bibinfo {volume} {126}},\ \bibinfo {pages} {157401} (\bibinfo {year} {2021})}\BibitemShut {NoStop}%
\bibitem [{\citenamefont {Yavorsky}\ \emph {et~al.}(2023)\citenamefont {Yavorsky}, \citenamefont {Kozhaev}, \citenamefont {Fedorov}, \citenamefont {Vikulin}, \citenamefont {Barshak}, \citenamefont {Berzhansky}, \citenamefont {Lyashko}, \citenamefont {Kapralov},\ and\ \citenamefont {Belotelov}}]{yav23}%
  \BibitemOpen
  \bibfield  {author} {\bibinfo {author} {\bibfnamefont {M.~A.}\ \bibnamefont {Yavorsky}}, \bibinfo {author} {\bibfnamefont {M.~A.}\ \bibnamefont {Kozhaev}}, \bibinfo {author} {\bibfnamefont {A.~Y.}\ \bibnamefont {Fedorov}}, \bibinfo {author} {\bibfnamefont {D.~V.}\ \bibnamefont {Vikulin}}, \bibinfo {author} {\bibfnamefont {E.~V.}\ \bibnamefont {Barshak}}, \bibinfo {author} {\bibfnamefont {V.~N.}\ \bibnamefont {Berzhansky}}, \bibinfo {author} {\bibfnamefont {S.~D.}\ \bibnamefont {Lyashko}}, \bibinfo {author} {\bibfnamefont {P.~O.}\ \bibnamefont {Kapralov}},\ and\ \bibinfo {author} {\bibfnamefont {V.~I.}\ \bibnamefont {Belotelov}},\ }\bibfield  {title} {\bibinfo {title} {{Topological Faraday Effect for Optical Vortices in Magnetic Films}},\ }\href {https://doi.org/10.1103/PhysRevLett.130.166901} {\bibfield  {journal} {\bibinfo  {journal} {Phys. Rev. Lett.}\ }\textbf {\bibinfo {volume} {130}},\ \bibinfo {pages} {166901} (\bibinfo {year} {2023})}\BibitemShut {NoStop}%
\bibitem [{\citenamefont {Gao}\ \emph {et~al.}(2023)\citenamefont {Gao}, \citenamefont {Prokhorenko}, \citenamefont {Nahas},\ and\ \citenamefont {Bellaiche}}]{gao23}%
  \BibitemOpen
  \bibfield  {author} {\bibinfo {author} {\bibfnamefont {L.}~\bibnamefont {Gao}}, \bibinfo {author} {\bibfnamefont {S.}~\bibnamefont {Prokhorenko}}, \bibinfo {author} {\bibfnamefont {Y.}~\bibnamefont {Nahas}},\ and\ \bibinfo {author} {\bibfnamefont {L.}~\bibnamefont {Bellaiche}},\ }\bibfield  {title} {\bibinfo {title} {Dynamical multiferroicity and magnetic topological structures induced by the orbital angular momentum of light in a nonmagnetic material},\ }\href {https://doi.org/10.1103/PhysRevLett.131.196801} {\bibfield  {journal} {\bibinfo  {journal} {Phys. Rev. Lett.}\ }\textbf {\bibinfo {volume} {131}},\ \bibinfo {pages} {196801} (\bibinfo {year} {2023})}\BibitemShut {NoStop}%
\bibitem [{\citenamefont {Gao}\ \emph {et~al.}(2024)\citenamefont {Gao}, \citenamefont {Prokhorenko}, \citenamefont {Nahas},\ and\ \citenamefont {Bellaiche}}]{gao24}%
  \BibitemOpen
  \bibfield  {author} {\bibinfo {author} {\bibfnamefont {L.}~\bibnamefont {Gao}}, \bibinfo {author} {\bibfnamefont {S.}~\bibnamefont {Prokhorenko}}, \bibinfo {author} {\bibfnamefont {Y.}~\bibnamefont {Nahas}},\ and\ \bibinfo {author} {\bibfnamefont {L.}~\bibnamefont {Bellaiche}},\ }\bibfield  {title} {\bibinfo {title} {{Dynamical Control of Topology in Polar Skyrmions via Twisted Light}},\ }\href {https://doi.org/10.1103/PhysRevLett.132.026902} {\bibfield  {journal} {\bibinfo  {journal} {Phys. Rev. Lett.}\ }\textbf {\bibinfo {volume} {132}},\ \bibinfo {pages} {026902} (\bibinfo {year} {2024})}\BibitemShut {NoStop}%
\bibitem [{\citenamefont {Manchon}\ \emph {et~al.}(2015)\citenamefont {Manchon}, \citenamefont {Koo}, \citenamefont {Nitta}, \citenamefont {Frolov},\ and\ \citenamefont {Duine}}]{man15}%
  \BibitemOpen
  \bibfield  {author} {\bibinfo {author} {\bibfnamefont {A.}~\bibnamefont {Manchon}}, \bibinfo {author} {\bibfnamefont {H.~C.}\ \bibnamefont {Koo}}, \bibinfo {author} {\bibfnamefont {J.}~\bibnamefont {Nitta}}, \bibinfo {author} {\bibfnamefont {S.~M.}\ \bibnamefont {Frolov}},\ and\ \bibinfo {author} {\bibfnamefont {R.~A.}\ \bibnamefont {Duine}},\ }\bibfield  {title} {\bibinfo {title} {{New perspectives for Rashba spin–orbit coupling}},\ }\href {https://doi.org/10.1038/nmat4360} {\bibfield  {journal} {\bibinfo  {journal} {Nat. Matter.}\ }\textbf {\bibinfo {volume} {14}},\ \bibinfo {pages} {871} (\bibinfo {year} {2015})}\BibitemShut {NoStop}%
\bibitem [{\citenamefont {Soumyanarayanan}\ \emph {et~al.}(2016)\citenamefont {Soumyanarayanan}, \citenamefont {Reyren}, \citenamefont {Fert},\ and\ \citenamefont {Panagopoulos}}]{sou16}%
  \BibitemOpen
  \bibfield  {author} {\bibinfo {author} {\bibfnamefont {A.}~\bibnamefont {Soumyanarayanan}}, \bibinfo {author} {\bibfnamefont {N.}~\bibnamefont {Reyren}}, \bibinfo {author} {\bibfnamefont {A.}~\bibnamefont {Fert}},\ and\ \bibinfo {author} {\bibfnamefont {C.}~\bibnamefont {Panagopoulos}},\ }\bibfield  {title} {\bibinfo {title} {{Emergent phenomena induced by spin–orbit coupling at surfaces and interfaces}},\ }\href {https://doi.org/10.1038/nature19820} {\bibfield  {journal} {\bibinfo  {journal} {Nature}\ }\textbf {\bibinfo {volume} {539}},\ \bibinfo {pages} {509} (\bibinfo {year} {2016})}\BibitemShut {NoStop}%
\bibitem [{\citenamefont {Bihlmayer}\ \emph {et~al.}(2022)\citenamefont {Bihlmayer}, \citenamefont {No\"{e}l}, \citenamefont {Vyalikh}, \citenamefont {Chulkov},\ and\ \citenamefont {Manchon}}]{bih22}%
  \BibitemOpen
  \bibfield  {author} {\bibinfo {author} {\bibfnamefont {G.}~\bibnamefont {Bihlmayer}}, \bibinfo {author} {\bibfnamefont {P.}~\bibnamefont {No\"{e}l}}, \bibinfo {author} {\bibfnamefont {D.~V.}\ \bibnamefont {Vyalikh}}, \bibinfo {author} {\bibfnamefont {E.~V.}\ \bibnamefont {Chulkov}},\ and\ \bibinfo {author} {\bibfnamefont {A.}~\bibnamefont {Manchon}},\ }\bibfield  {title} {\bibinfo {title} {{Rashba-like physics in condensed matter}},\ }\href {https://doi.org/10.1038/s42254-022-00490-y} {\bibfield  {journal} {\bibinfo  {journal} {Nat. Rev. Phys.}\ }\textbf {\bibinfo {volume} {4}},\ \bibinfo {pages} {642} (\bibinfo {year} {2022})}\BibitemShut {NoStop}%
\bibitem [{\citenamefont {Dresselhaus}(1955)}]{dre55}%
  \BibitemOpen
  \bibfield  {author} {\bibinfo {author} {\bibfnamefont {G.}~\bibnamefont {Dresselhaus}},\ }\bibfield  {title} {\bibinfo {title} {{Spin-Orbit Coupling Effects in Zinc Blende Structures}},\ }\href {https://doi.org/10.1103/PhysRev.100.580} {\bibfield  {journal} {\bibinfo  {journal} {Phys. Rev.}\ }\textbf {\bibinfo {volume} {100}},\ \bibinfo {pages} {580} (\bibinfo {year} {1955})}\BibitemShut {NoStop}%
\bibitem [{\citenamefont {Bychkov}\ and\ \citenamefont {Rashba}(1984)}]{byc84}%
  \BibitemOpen
  \bibfield  {author} {\bibinfo {author} {\bibfnamefont {Y.~A.}\ \bibnamefont {Bychkov}}\ and\ \bibinfo {author} {\bibfnamefont {E.~I.}\ \bibnamefont {Rashba}},\ }\bibfield  {title} {\bibinfo {title} {{Oscillatory effects and the magnetic susceptibility of carriers in inversion layers}},\ }\href {https://doi.org/10.1088/0022-3719/17/33/015} {\bibfield  {journal} {\bibinfo  {journal} {J. Phys. C}\ }\textbf {\bibinfo {volume} {17}},\ \bibinfo {pages} {6039} (\bibinfo {year} {1984})}\BibitemShut {NoStop}%
\bibitem [{\citenamefont {Aronov}\ and\ \citenamefont {Lyanda-Geller}(1989)}]{aro89}%
  \BibitemOpen
  \bibfield  {author} {\bibinfo {author} {\bibfnamefont {A.~G.}\ \bibnamefont {Aronov}}\ and\ \bibinfo {author} {\bibfnamefont {Y.~B.}\ \bibnamefont {Lyanda-Geller}},\ }\bibfield  {title} {\bibinfo {title} {{Nuclear electric resonance and orientation of carrier spins by an electric field}},\ }\href@noop {} {\bibfield  {journal} {\bibinfo  {journal} {JETP Lett.}\ }\textbf {\bibinfo {volume} {50}},\ \bibinfo {pages} {431} (\bibinfo {year} {1989})},\ \bibinfo {note} {{Pis'ma Zh. Eksp. Teor. Fiz. {\bf 50}, 398 (1989)}}\BibitemShut {NoStop}%
\bibitem [{\citenamefont {Edelstein}(1990)}]{edelstein_1990}%
  \BibitemOpen
  \bibfield  {author} {\bibinfo {author} {\bibfnamefont {V.~M.}\ \bibnamefont {Edelstein}},\ }\bibfield  {title} {\bibinfo {title} {{Spin polarization of conduction electrons induced by electric current in two-dimensional asymmetric electron systems}},\ }\href {https://doi.org/10.1016/0038-1098(90)90963-C} {\bibfield  {journal} {\bibinfo  {journal} {Solid State Commun.}\ }\textbf {\bibinfo {volume} {73}},\ \bibinfo {pages} {233} (\bibinfo {year} {1990})}\BibitemShut {NoStop}%
\bibitem [{\citenamefont {Aronov}\ \emph {et~al.}(1991)\citenamefont {Aronov}, \citenamefont {Lyanda-Geller},\ and\ \citenamefont {Pikus}}]{aro91}%
  \BibitemOpen
  \bibfield  {author} {\bibinfo {author} {\bibfnamefont {A.~G.}\ \bibnamefont {Aronov}}, \bibinfo {author} {\bibfnamefont {Y.~B.}\ \bibnamefont {Lyanda-Geller}},\ and\ \bibinfo {author} {\bibfnamefont {G.~E.}\ \bibnamefont {Pikus}},\ }\bibfield  {title} {\bibinfo {title} {{Spin polarization of electrons by an electric current}},\ }\href@noop {} {\bibfield  {journal} {\bibinfo  {journal} {Sov. Phys. JETP}\ }\textbf {\bibinfo {volume} {73}},\ \bibinfo {pages} {537} (\bibinfo {year} {1991})},\ \bibinfo {note} {{Zh. Eksp. Teor. Fiz. {\bf 100}, 973 (1991)}}\BibitemShut {NoStop}%
\bibitem [{\citenamefont {Mishchenko}\ \emph {et~al.}(2004)\citenamefont {Mishchenko}, \citenamefont {Shytov},\ and\ \citenamefont {Halperin}}]{mis04}%
  \BibitemOpen
  \bibfield  {author} {\bibinfo {author} {\bibfnamefont {E.~G.}\ \bibnamefont {Mishchenko}}, \bibinfo {author} {\bibfnamefont {A.~V.}\ \bibnamefont {Shytov}},\ and\ \bibinfo {author} {\bibfnamefont {B.~I.}\ \bibnamefont {Halperin}},\ }\bibfield  {title} {\bibinfo {title} {{Spin Current and Polarization in Impure Two-Dimensional Electron Systems with Spin-Orbit Coupling}},\ }\href {https://doi.org/10.1103/PhysRevLett.93.226602} {\bibfield  {journal} {\bibinfo  {journal} {Phys. Rev. Lett.}\ }\textbf {\bibinfo {volume} {93}},\ \bibinfo {pages} {226602} (\bibinfo {year} {2004})}\BibitemShut {NoStop}%
\bibitem [{\citenamefont {\ifmmode \check{Z}\else \v{Z}\fi{}uti\ifmmode~\acute{c}\else \'{c}\fi{}}\ \emph {et~al.}(2004)\citenamefont {\ifmmode \check{Z}\else \v{Z}\fi{}uti\ifmmode~\acute{c}\else \'{c}\fi{}}, \citenamefont {Fabian},\ and\ \citenamefont {Das~Sarma}}]{sar04}%
  \BibitemOpen
  \bibfield  {author} {\bibinfo {author} {\bibfnamefont {I.}~\bibnamefont {\ifmmode \check{Z}\else \v{Z}\fi{}uti\ifmmode~\acute{c}\else \'{c}\fi{}}}, \bibinfo {author} {\bibfnamefont {J.}~\bibnamefont {Fabian}},\ and\ \bibinfo {author} {\bibfnamefont {S.}~\bibnamefont {Das~Sarma}},\ }\bibfield  {title} {\bibinfo {title} {{Spintronics: Fundamentals and applications}},\ }\href {https://doi.org/10.1103/RevModPhys.76.323} {\bibfield  {journal} {\bibinfo  {journal} {Rev. Mod. Phys.}\ }\textbf {\bibinfo {volume} {76}},\ \bibinfo {pages} {323} (\bibinfo {year} {2004})}\BibitemShut {NoStop}%
\bibitem [{\citenamefont {Ando}\ and\ \citenamefont {Shiraishi}(2017)}]{and17}%
  \BibitemOpen
  \bibfield  {author} {\bibinfo {author} {\bibfnamefont {Y.}~\bibnamefont {Ando}}\ and\ \bibinfo {author} {\bibfnamefont {M.}~\bibnamefont {Shiraishi}},\ }\bibfield  {title} {\bibinfo {title} {{Spin to Charge Interconversion Phenomena in the Interface and Surface States}},\ }\href {https://doi.org/10.7566/JPSJ.86.011001} {\bibfield  {journal} {\bibinfo  {journal} {J. Phys. Soc. Jpn.}\ }\textbf {\bibinfo {volume} {86}},\ \bibinfo {pages} {011001} (\bibinfo {year} {2017})}\BibitemShut {NoStop}%
\bibitem [{\citenamefont {Feng}\ \emph {et~al.}(2017)\citenamefont {Feng}, \citenamefont {Shen}, \citenamefont {Yang}, \citenamefont {Wang}, \citenamefont {Zeng}, \citenamefont {Wu}, \citenamefont {Chintalapati},\ and\ \citenamefont {Chang}}]{fen17}%
  \BibitemOpen
  \bibfield  {author} {\bibinfo {author} {\bibfnamefont {Y.~P.}\ \bibnamefont {Feng}}, \bibinfo {author} {\bibfnamefont {L.}~\bibnamefont {Shen}}, \bibinfo {author} {\bibfnamefont {M.}~\bibnamefont {Yang}}, \bibinfo {author} {\bibfnamefont {A.}~\bibnamefont {Wang}}, \bibinfo {author} {\bibfnamefont {M.}~\bibnamefont {Zeng}}, \bibinfo {author} {\bibfnamefont {Q.}~\bibnamefont {Wu}}, \bibinfo {author} {\bibfnamefont {S.}~\bibnamefont {Chintalapati}},\ and\ \bibinfo {author} {\bibfnamefont {C.-R.}\ \bibnamefont {Chang}},\ }\bibfield  {title} {\bibinfo {title} {{Prospects of spintronics based on 2D materials}},\ }\href {https://doi.org/https://doi.org/10.1002/wcms.1313} {\bibfield  {journal} {\bibinfo  {journal} {WIREs Computational Molecular Science}\ }\textbf {\bibinfo {volume} {7}},\ \bibinfo {pages} {e1313} (\bibinfo {year} {2017})}\BibitemShut {NoStop}%
\bibitem [{\citenamefont {Ganichev}\ and\ \citenamefont {Prettl}(2003)}]{gan03}%
  \BibitemOpen
  \bibfield  {author} {\bibinfo {author} {\bibfnamefont {S.~D.}\ \bibnamefont {Ganichev}}\ and\ \bibinfo {author} {\bibfnamefont {W.}~\bibnamefont {Prettl}},\ }\bibfield  {title} {\bibinfo {title} {{Spin photocurrents in quantum wells}},\ }\href {https://doi.org/10.1088/0953-8984/15/20/204} {\bibfield  {journal} {\bibinfo  {journal} {J. Phys. Condens. Matter}\ }\textbf {\bibinfo {volume} {15}},\ \bibinfo {pages} {R935} (\bibinfo {year} {2003})}\BibitemShut {NoStop}%
\bibitem [{\citenamefont {Diehl}\ \emph {et~al.}(2007)\citenamefont {Diehl}, \citenamefont {Shalygin}, \citenamefont {Bel'kov}, \citenamefont {Hoffmann}, \citenamefont {Danilov}, \citenamefont {Herrle}, \citenamefont {Tarasenko}, \citenamefont {Schuh}, \citenamefont {Gerl}, \citenamefont {Wegscheider}, \citenamefont {Prettl},\ and\ \citenamefont {Ganichev}}]{die07}%
  \BibitemOpen
  \bibfield  {author} {\bibinfo {author} {\bibfnamefont {H.}~\bibnamefont {Diehl}}, \bibinfo {author} {\bibfnamefont {V.~A.}\ \bibnamefont {Shalygin}}, \bibinfo {author} {\bibfnamefont {V.~V.}\ \bibnamefont {Bel'kov}}, \bibinfo {author} {\bibfnamefont {C.}~\bibnamefont {Hoffmann}}, \bibinfo {author} {\bibfnamefont {S.~N.}\ \bibnamefont {Danilov}}, \bibinfo {author} {\bibfnamefont {T.}~\bibnamefont {Herrle}}, \bibinfo {author} {\bibfnamefont {S.~A.}\ \bibnamefont {Tarasenko}}, \bibinfo {author} {\bibfnamefont {D.}~\bibnamefont {Schuh}}, \bibinfo {author} {\bibfnamefont {C.}~\bibnamefont {Gerl}}, \bibinfo {author} {\bibfnamefont {W.}~\bibnamefont {Wegscheider}}, \bibinfo {author} {\bibfnamefont {W.}~\bibnamefont {Prettl}},\ and\ \bibinfo {author} {\bibfnamefont {S.~D.}\ \bibnamefont {Ganichev}},\ }\bibfield  {title} {\bibinfo {title} {{Spin photocurrents in (110)-grown quantum well structures}},\ }\href {https://doi.org/10.1088/1367-2630/9/9/349} {\bibfield  {journal} {\bibinfo  {journal} {New J. Phys.}\
  }\textbf {\bibinfo {volume} {9}},\ \bibinfo {pages} {349} (\bibinfo {year} {2007})}\BibitemShut {NoStop}%
\bibitem [{\citenamefont {Moore}\ and\ \citenamefont {Orenstein}(2010)}]{moo10}%
  \BibitemOpen
  \bibfield  {author} {\bibinfo {author} {\bibfnamefont {J.~E.}\ \bibnamefont {Moore}}\ and\ \bibinfo {author} {\bibfnamefont {J.}~\bibnamefont {Orenstein}},\ }\bibfield  {title} {\bibinfo {title} {{Confinement-Induced Berry Phase and Helicity-Dependent Photocurrents}},\ }\href {https://doi.org/10.1103/PhysRevLett.105.026805} {\bibfield  {journal} {\bibinfo  {journal} {Phys. Rev. Lett.}\ }\textbf {\bibinfo {volume} {105}},\ \bibinfo {pages} {026805} (\bibinfo {year} {2010})}\BibitemShut {NoStop}%
\bibitem [{\citenamefont {McIver}\ \emph {et~al.}(2012)\citenamefont {McIver}, \citenamefont {Hsieh}, \citenamefont {Steinberg}, \citenamefont {Jarillo-Herrero},\ and\ \citenamefont {Gedik}}]{mci12}%
  \BibitemOpen
  \bibfield  {author} {\bibinfo {author} {\bibfnamefont {J.~W.}\ \bibnamefont {McIver}}, \bibinfo {author} {\bibfnamefont {D.}~\bibnamefont {Hsieh}}, \bibinfo {author} {\bibfnamefont {H.}~\bibnamefont {Steinberg}}, \bibinfo {author} {\bibfnamefont {P.}~\bibnamefont {Jarillo-Herrero}},\ and\ \bibinfo {author} {\bibfnamefont {N.}~\bibnamefont {Gedik}},\ }\bibfield  {title} {\bibinfo {title} {{Control over topological insulator photocurrents with light polarization}},\ }\href {https://doi.org/10.1038/nnano.2011.214} {\bibfield  {journal} {\bibinfo  {journal} {Nat. Nanotechnol}\ }\textbf {\bibinfo {volume} {7}},\ \bibinfo {pages} {96} (\bibinfo {year} {2012})}\BibitemShut {NoStop}%
\bibitem [{\citenamefont {Gor'kov}\ and\ \citenamefont {Rashba}(2001)}]{gor01}%
  \BibitemOpen
  \bibfield  {author} {\bibinfo {author} {\bibfnamefont {L.~P.}\ \bibnamefont {Gor'kov}}\ and\ \bibinfo {author} {\bibfnamefont {E.~I.}\ \bibnamefont {Rashba}},\ }\bibfield  {title} {\bibinfo {title} {{Superconducting 2D System with Lifted Spin Degeneracy: Mixed Singlet-Triplet State}},\ }\href {https://doi.org/10.1103/PhysRevLett.87.037004} {\bibfield  {journal} {\bibinfo  {journal} {Phys. Rev. Lett.}\ }\textbf {\bibinfo {volume} {87}},\ \bibinfo {pages} {037004} (\bibinfo {year} {2001})}\BibitemShut {NoStop}%
\bibitem [{\citenamefont {Sato}\ \emph {et~al.}(2009)\citenamefont {Sato}, \citenamefont {Takahashi},\ and\ \citenamefont {Fujimoto}}]{sat09-1}%
  \BibitemOpen
  \bibfield  {author} {\bibinfo {author} {\bibfnamefont {M.}~\bibnamefont {Sato}}, \bibinfo {author} {\bibfnamefont {Y.}~\bibnamefont {Takahashi}},\ and\ \bibinfo {author} {\bibfnamefont {S.}~\bibnamefont {Fujimoto}},\ }\bibfield  {title} {\bibinfo {title} {{Non-Abelian Topological Order in $s$-Wave Superfluids of Ultracold Fermionic Atoms}},\ }\href {https://doi.org/10.1103/PhysRevLett.103.020401} {\bibfield  {journal} {\bibinfo  {journal} {Phys. Rev. Lett.}\ }\textbf {\bibinfo {volume} {103}},\ \bibinfo {pages} {020401} (\bibinfo {year} {2009})}\BibitemShut {NoStop}%
\bibitem [{\citenamefont {Sato}\ and\ \citenamefont {Fujimoto}(2009)}]{sat09-2}%
  \BibitemOpen
  \bibfield  {author} {\bibinfo {author} {\bibfnamefont {M.}~\bibnamefont {Sato}}\ and\ \bibinfo {author} {\bibfnamefont {S.}~\bibnamefont {Fujimoto}},\ }\bibfield  {title} {\bibinfo {title} {{Topological phases of noncentrosymmetric superconductors: Edge states, Majorana fermions, and non-Abelian statistics}},\ }\href {https://doi.org/10.1103/PhysRevB.79.094504} {\bibfield  {journal} {\bibinfo  {journal} {Phys. Rev. B}\ }\textbf {\bibinfo {volume} {79}},\ \bibinfo {pages} {094504} (\bibinfo {year} {2009})}\BibitemShut {NoStop}%
\bibitem [{\citenamefont {Sato}\ \emph {et~al.}(2010)\citenamefont {Sato}, \citenamefont {Takahashi},\ and\ \citenamefont {Fujimoto}}]{sat10}%
  \BibitemOpen
  \bibfield  {author} {\bibinfo {author} {\bibfnamefont {M.}~\bibnamefont {Sato}}, \bibinfo {author} {\bibfnamefont {Y.}~\bibnamefont {Takahashi}},\ and\ \bibinfo {author} {\bibfnamefont {S.}~\bibnamefont {Fujimoto}},\ }\bibfield  {title} {\bibinfo {title} {{Non-Abelian topological orders and Majorana fermions in spin-singlet superconductors}},\ }\href {https://doi.org/10.1103/PhysRevB.82.134521} {\bibfield  {journal} {\bibinfo  {journal} {Phys. Rev. B}\ }\textbf {\bibinfo {volume} {82}},\ \bibinfo {pages} {134521} (\bibinfo {year} {2010})}\BibitemShut {NoStop}%
\bibitem [{\citenamefont {Lutchyn}\ \emph {et~al.}(2010)\citenamefont {Lutchyn}, \citenamefont {Sau},\ and\ \citenamefont {Das~Sarma}}]{lut10}%
  \BibitemOpen
  \bibfield  {author} {\bibinfo {author} {\bibfnamefont {R.~M.}\ \bibnamefont {Lutchyn}}, \bibinfo {author} {\bibfnamefont {J.~D.}\ \bibnamefont {Sau}},\ and\ \bibinfo {author} {\bibfnamefont {S.}~\bibnamefont {Das~Sarma}},\ }\bibfield  {title} {\bibinfo {title} {Majorana fermions and a topological phase transition in semiconductor-superconductor heterostructures},\ }\href {https://doi.org/10.1103/PhysRevLett.105.077001} {\bibfield  {journal} {\bibinfo  {journal} {Phys. Rev. Lett.}\ }\textbf {\bibinfo {volume} {105}},\ \bibinfo {pages} {077001} (\bibinfo {year} {2010})}\BibitemShut {NoStop}%
\bibitem [{\citenamefont {Sau}\ \emph {et~al.}(2010)\citenamefont {Sau}, \citenamefont {Lutchyn}, \citenamefont {Tewari},\ and\ \citenamefont {Das~Sarma}}]{sau10}%
  \BibitemOpen
  \bibfield  {author} {\bibinfo {author} {\bibfnamefont {J.~D.}\ \bibnamefont {Sau}}, \bibinfo {author} {\bibfnamefont {R.~M.}\ \bibnamefont {Lutchyn}}, \bibinfo {author} {\bibfnamefont {S.}~\bibnamefont {Tewari}},\ and\ \bibinfo {author} {\bibfnamefont {S.}~\bibnamefont {Das~Sarma}},\ }\bibfield  {title} {\bibinfo {title} {{Generic New Platform for Topological Quantum Computation Using Semiconductor Heterostructures}},\ }\href {https://doi.org/10.1103/PhysRevLett.104.040502} {\bibfield  {journal} {\bibinfo  {journal} {Phys. Rev. Lett.}\ }\textbf {\bibinfo {volume} {104}},\ \bibinfo {pages} {040502} (\bibinfo {year} {2010})}\BibitemShut {NoStop}%
\bibitem [{\citenamefont {Alicea}(2010)}]{ali10}%
  \BibitemOpen
  \bibfield  {author} {\bibinfo {author} {\bibfnamefont {J.}~\bibnamefont {Alicea}},\ }\bibfield  {title} {\bibinfo {title} {{Majorana fermions in a tunable semiconductor device}},\ }\href {https://doi.org/10.1103/PhysRevB.81.125318} {\bibfield  {journal} {\bibinfo  {journal} {Phys. Rev. B}\ }\textbf {\bibinfo {volume} {81}},\ \bibinfo {pages} {125318} (\bibinfo {year} {2010})}\BibitemShut {NoStop}%
\bibitem [{\citenamefont {Bauer}\ and\ \citenamefont {Sigrist}(2012)}]{bau}%
  \BibitemOpen
  \bibfield  {author} {\bibinfo {author} {\bibfnamefont {E.}~\bibnamefont {Bauer}}\ and\ \bibinfo {author} {\bibfnamefont {M.}~\bibnamefont {Sigrist}},\ }\href@noop {} {\emph {\bibinfo {title} {{Non-Centrosymmetric Superconductors}}}},\ Lecture Notes in Physics\ (\bibinfo  {publisher} {Springer-Verlag, Berlin},\ \bibinfo {year} {2012})\BibitemShut {NoStop}%
\bibitem [{\citenamefont {Yip}(2014)}]{yip14}%
  \BibitemOpen
  \bibfield  {author} {\bibinfo {author} {\bibfnamefont {S.}~\bibnamefont {Yip}},\ }\bibfield  {title} {\bibinfo {title} {{Noncentrosymmetric Superconductors}},\ }\href {https://doi.org/https://doi.org/10.1146/annurev-conmatphys-031113-133912} {\bibfield  {journal} {\bibinfo  {journal} {Annu. Rev. Condens. Matter Phys.}\ }\textbf {\bibinfo {volume} {5}},\ \bibinfo {pages} {15} (\bibinfo {year} {2014})}\BibitemShut {NoStop}%
\bibitem [{\citenamefont {Zhan}(2009)}]{zha09}%
  \BibitemOpen
  \bibfield  {author} {\bibinfo {author} {\bibfnamefont {Q.}~\bibnamefont {Zhan}},\ }\bibfield  {title} {\bibinfo {title} {{Cylindrical vector beams: from mathematical concepts to applications}},\ }\href {https://doi.org/10.1364/AOP.1.000001} {\bibfield  {journal} {\bibinfo  {journal} {Adv. Opt. Photon.}\ }\textbf {\bibinfo {volume} {1}},\ \bibinfo {pages} {1} (\bibinfo {year} {2009})}\BibitemShut {NoStop}%
\bibitem [{\citenamefont {Shen}\ \emph {et~al.}(2019)\citenamefont {Shen}, \citenamefont {Wang}, \citenamefont {Xie}, \citenamefont {Min}, \citenamefont {Fu}, \citenamefont {Liu}, \citenamefont {Gong},\ and\ \citenamefont {Yuan}}]{she19}%
  \BibitemOpen
  \bibfield  {author} {\bibinfo {author} {\bibfnamefont {Y.}~\bibnamefont {Shen}}, \bibinfo {author} {\bibfnamefont {X.}~\bibnamefont {Wang}}, \bibinfo {author} {\bibfnamefont {Z.}~\bibnamefont {Xie}}, \bibinfo {author} {\bibfnamefont {C.}~\bibnamefont {Min}}, \bibinfo {author} {\bibfnamefont {X.}~\bibnamefont {Fu}}, \bibinfo {author} {\bibfnamefont {Q.}~\bibnamefont {Liu}}, \bibinfo {author} {\bibfnamefont {M.}~\bibnamefont {Gong}},\ and\ \bibinfo {author} {\bibfnamefont {X.}~\bibnamefont {Yuan}},\ }\bibfield  {title} {\bibinfo {title} {{Optical vortices 30 years on: OAM manipulation from topological charge to multiple singularities}},\ }\href {https://doi.org/10.1038/s41377-019-0194-2} {\bibfield  {journal} {\bibinfo  {journal} {Light Sci. Appl.}\ }\textbf {\bibinfo {volume} {8}},\ \bibinfo {pages} {90} (\bibinfo {year} {2019})}\BibitemShut {NoStop}%
\bibitem [{\citenamefont {Forbes}(2019)}]{for19}%
  \BibitemOpen
  \bibfield  {author} {\bibinfo {author} {\bibfnamefont {A.}~\bibnamefont {Forbes}},\ }\bibfield  {title} {\bibinfo {title} {{Structured Light from Lasers}},\ }\href {https://doi.org/https://doi.org/10.1002/lpor.201900140} {\bibfield  {journal} {\bibinfo  {journal} {Laser Photonics Rev.}\ }\textbf {\bibinfo {volume} {13}},\ \bibinfo {pages} {1900140} (\bibinfo {year} {2019})}\BibitemShut {NoStop}%
\bibitem [{\citenamefont {Nape}\ \emph {et~al.}(2023)\citenamefont {Nape}, \citenamefont {Sephton}, \citenamefont {Ornelas}, \citenamefont {Moodley},\ and\ \citenamefont {Forbes}}]{nap23}%
  \BibitemOpen
  \bibfield  {author} {\bibinfo {author} {\bibfnamefont {I.}~\bibnamefont {Nape}}, \bibinfo {author} {\bibfnamefont {B.}~\bibnamefont {Sephton}}, \bibinfo {author} {\bibfnamefont {P.}~\bibnamefont {Ornelas}}, \bibinfo {author} {\bibfnamefont {C.}~\bibnamefont {Moodley}},\ and\ \bibinfo {author} {\bibfnamefont {A.}~\bibnamefont {Forbes}},\ }\bibfield  {title} {\bibinfo {title} {{Quantum structured light in high dimensions}},\ }\href {https://doi.org/10.1063/5.0138224} {\bibfield  {journal} {\bibinfo  {journal} {APL Photonics}\ }\textbf {\bibinfo {volume} {8}},\ \bibinfo {pages} {051101} (\bibinfo {year} {2023})}\BibitemShut {NoStop}%
\bibitem [{\citenamefont {Clayburn}\ \emph {et~al.}(2013)\citenamefont {Clayburn}, \citenamefont {McCarter}, \citenamefont {Dreiling}, \citenamefont {Poelker}, \citenamefont {Ryan},\ and\ \citenamefont {Gay}}]{cla13}%
  \BibitemOpen
  \bibfield  {author} {\bibinfo {author} {\bibfnamefont {N.~B.}\ \bibnamefont {Clayburn}}, \bibinfo {author} {\bibfnamefont {J.~L.}\ \bibnamefont {McCarter}}, \bibinfo {author} {\bibfnamefont {J.~M.}\ \bibnamefont {Dreiling}}, \bibinfo {author} {\bibfnamefont {M.}~\bibnamefont {Poelker}}, \bibinfo {author} {\bibfnamefont {D.~M.}\ \bibnamefont {Ryan}},\ and\ \bibinfo {author} {\bibfnamefont {T.~J.}\ \bibnamefont {Gay}},\ }\bibfield  {title} {\bibinfo {title} {{Search for spin-polarized photoemission from GaAs using light with orbital angular momentum}},\ }\href {https://doi.org/10.1103/PhysRevB.87.035204} {\bibfield  {journal} {\bibinfo  {journal} {Phys. Rev. B}\ }\textbf {\bibinfo {volume} {87}},\ \bibinfo {pages} {035204} (\bibinfo {year} {2013})}\BibitemShut {NoStop}%
\bibitem [{\citenamefont {Sordillo}\ \emph {et~al.}(2019)\citenamefont {Sordillo}, \citenamefont {Mamani}, \citenamefont {Sharonov},\ and\ \citenamefont {Alfano}}]{sor19}%
  \BibitemOpen
  \bibfield  {author} {\bibinfo {author} {\bibfnamefont {L.~A.}\ \bibnamefont {Sordillo}}, \bibinfo {author} {\bibfnamefont {S.}~\bibnamefont {Mamani}}, \bibinfo {author} {\bibfnamefont {M.}~\bibnamefont {Sharonov}},\ and\ \bibinfo {author} {\bibfnamefont {R.~R.}\ \bibnamefont {Alfano}},\ }\bibfield  {title} {\bibinfo {title} {{The interaction of twisted Laguerre-Gaussian light with a GaAs photocathode to investigate photogenerated polarized electrons}},\ }\href {https://doi.org/10.1063/1.5078503} {\bibfield  {journal} {\bibinfo  {journal} {Appl. Phys. Lett.}\ }\textbf {\bibinfo {volume} {114}},\ \bibinfo {pages} {041104} (\bibinfo {year} {2019})}\BibitemShut {NoStop}%
\bibitem [{\citenamefont {Ji}\ \emph {et~al.}(2020)\citenamefont {Ji}, \citenamefont {Liu}, \citenamefont {Krylyuk}, \citenamefont {Fan}, \citenamefont {Zhang}, \citenamefont {Pan}, \citenamefont {Feng}, \citenamefont {Davydov},\ and\ \citenamefont {Agarwal}}]{ji20}%
  \BibitemOpen
  \bibfield  {author} {\bibinfo {author} {\bibfnamefont {Z.}~\bibnamefont {Ji}}, \bibinfo {author} {\bibfnamefont {W.}~\bibnamefont {Liu}}, \bibinfo {author} {\bibfnamefont {S.}~\bibnamefont {Krylyuk}}, \bibinfo {author} {\bibfnamefont {X.}~\bibnamefont {Fan}}, \bibinfo {author} {\bibfnamefont {Z.}~\bibnamefont {Zhang}}, \bibinfo {author} {\bibfnamefont {A.}~\bibnamefont {Pan}}, \bibinfo {author} {\bibfnamefont {L.}~\bibnamefont {Feng}}, \bibinfo {author} {\bibfnamefont {A.}~\bibnamefont {Davydov}},\ and\ \bibinfo {author} {\bibfnamefont {R.}~\bibnamefont {Agarwal}},\ }\bibfield  {title} {\bibinfo {title} {Photocurrent detection of the orbital angular momentum of light},\ }\href {https://doi.org/10.1126/science.aba9192} {\bibfield  {journal} {\bibinfo  {journal} {Science}\ }\textbf {\bibinfo {volume} {368}},\ \bibinfo {pages} {763} (\bibinfo {year} {2020})}\BibitemShut {NoStop}%
\bibitem [{\citenamefont {Ishihara}\ \emph {et~al.}(2023)\citenamefont {Ishihara}, \citenamefont {Mori}, \citenamefont {Suzuki}, \citenamefont {Sato}, \citenamefont {Morita}, \citenamefont {Kohda}, \citenamefont {Ohno},\ and\ \citenamefont {Miyajima}}]{Ishihara_2023}%
  \BibitemOpen
  \bibfield  {author} {\bibinfo {author} {\bibfnamefont {J.}~\bibnamefont {Ishihara}}, \bibinfo {author} {\bibfnamefont {T.}~\bibnamefont {Mori}}, \bibinfo {author} {\bibfnamefont {T.}~\bibnamefont {Suzuki}}, \bibinfo {author} {\bibfnamefont {S.}~\bibnamefont {Sato}}, \bibinfo {author} {\bibfnamefont {K.}~\bibnamefont {Morita}}, \bibinfo {author} {\bibfnamefont {M.}~\bibnamefont {Kohda}}, \bibinfo {author} {\bibfnamefont {Y.}~\bibnamefont {Ohno}},\ and\ \bibinfo {author} {\bibfnamefont {K.}~\bibnamefont {Miyajima}},\ }\bibfield  {title} {\bibinfo {title} {{Imprinting spatial helicity structure of vector vortex beam on spin texture in semiconductors}},\ }\href {https://doi.org/10.1103/PhysRevLett.130.126701} {\bibfield  {journal} {\bibinfo  {journal} {Phys. Rev. Lett.}\ }\textbf {\bibinfo {volume} {130}},\ \bibinfo {pages} {126701} (\bibinfo {year} {2023})}\BibitemShut {NoStop}%
\bibitem [{\citenamefont {Matsumoto}\ \emph {et~al.}(2024)\citenamefont {Matsumoto}, \citenamefont {Sato}, \citenamefont {Akei}, \citenamefont {Nakano}, \citenamefont {Iba}, \citenamefont {Ishihara}, \citenamefont {Miyamoto}, \citenamefont {Yokoshi}, \citenamefont {Omatsu},\ and\ \citenamefont {Morita}}]{mat24}%
  \BibitemOpen
  \bibfield  {author} {\bibinfo {author} {\bibfnamefont {T.}~\bibnamefont {Matsumoto}}, \bibinfo {author} {\bibfnamefont {S.}~\bibnamefont {Sato}}, \bibinfo {author} {\bibfnamefont {S.}~\bibnamefont {Akei}}, \bibinfo {author} {\bibfnamefont {Y.}~\bibnamefont {Nakano}}, \bibinfo {author} {\bibfnamefont {S.}~\bibnamefont {Iba}}, \bibinfo {author} {\bibfnamefont {J.}~\bibnamefont {Ishihara}}, \bibinfo {author} {\bibfnamefont {K.}~\bibnamefont {Miyamoto}}, \bibinfo {author} {\bibfnamefont {N.}~\bibnamefont {Yokoshi}}, \bibinfo {author} {\bibfnamefont {T.}~\bibnamefont {Omatsu}},\ and\ \bibinfo {author} {\bibfnamefont {K.}~\bibnamefont {Morita}},\ }\bibfield  {title} {\bibinfo {title} {{Coherent transfer of the higher-order polarization state of photons to the spin structure state of electrons in a semiconductor}},\ }\href {https://doi.org/10.1364/OPTICAQ.527615} {\bibfield  {journal} {\bibinfo  {journal} {Optica Quantum}\ }\textbf {\bibinfo {volume} {2}},\ \bibinfo {pages} {245} (\bibinfo {year}
  {2024})}\BibitemShut {NoStop}%
\bibitem [{\citenamefont {Terashima}\ \emph {et~al.}(2025)\citenamefont {Terashima}, \citenamefont {Ito}, \citenamefont {Kakue}, \citenamefont {Yokoshi},\ and\ \citenamefont {Morita}}]{ter25}%
  \BibitemOpen
  \bibfield  {author} {\bibinfo {author} {\bibfnamefont {K.}~\bibnamefont {Terashima}}, \bibinfo {author} {\bibfnamefont {R.}~\bibnamefont {Ito}}, \bibinfo {author} {\bibfnamefont {T.}~\bibnamefont {Kakue}}, \bibinfo {author} {\bibfnamefont {N.}~\bibnamefont {Yokoshi}},\ and\ \bibinfo {author} {\bibfnamefont {K.}~\bibnamefont {Morita}},\ }\bibfield  {title} {\bibinfo {title} {{Optical initialization and manipulation of higher-order electron states with spin and orbital angular momentum in a semiconductor quantum disk}},\ }\href {https://doi.org/10.1364/OE.550223} {\bibfield  {journal} {\bibinfo  {journal} {Opt. Express}\ }\textbf {\bibinfo {volume} {33}},\ \bibinfo {pages} {4700} (\bibinfo {year} {2025})}\BibitemShut {NoStop}%
\bibitem [{\citenamefont {Levitov}\ \emph {et~al.}(1985)\citenamefont {Levitov}, \citenamefont {Nazarov},\ and\ \citenamefont {Eliashberg}}]{levitov_1985}%
  \BibitemOpen
  \bibfield  {author} {\bibinfo {author} {\bibfnamefont {L.}~\bibnamefont {Levitov}}, \bibinfo {author} {\bibfnamefont {Y.}~\bibnamefont {Nazarov}},\ and\ \bibinfo {author} {\bibfnamefont {G.}~\bibnamefont {Eliashberg}},\ }\bibfield  {title} {\bibinfo {title} {{Magnetoelectric effects in conductors with mirror isomer symmetry}},\ }\href@noop {} {\bibfield  {journal} {\bibinfo  {journal} {Sov. Phys. JETP}\ }\textbf {\bibinfo {volume} {61}},\ \bibinfo {pages} {133} (\bibinfo {year} {1985})}\BibitemShut {NoStop}%
\bibitem [{\citenamefont {Fujimoto}(2005)}]{fuj05}%
  \BibitemOpen
  \bibfield  {author} {\bibinfo {author} {\bibfnamefont {S.}~\bibnamefont {Fujimoto}},\ }\bibfield  {title} {\bibinfo {title} {{Magnetoelectric effects in heavy-fermion superconductors without inversion symmetry}},\ }\href {https://doi.org/10.1103/PhysRevB.72.024515} {\bibfield  {journal} {\bibinfo  {journal} {Phys. Rev. B}\ }\textbf {\bibinfo {volume} {72}},\ \bibinfo {pages} {024515} (\bibinfo {year} {2005})}\BibitemShut {NoStop}%
\bibitem [{\citenamefont {Fujimoto}(2007)}]{fujimoto_2007}%
  \BibitemOpen
  \bibfield  {author} {\bibinfo {author} {\bibfnamefont {S.}~\bibnamefont {Fujimoto}},\ }\bibfield  {title} {\bibinfo {title} {{Fermi liquid theory for heavy fermion superconductors without inversion symmetry: Magnetism and transport coefficients}},\ }\href {https://doi.org/10.1143/JPSJ.76.034712} {\bibfield  {journal} {\bibinfo  {journal} {J. Phys. Soc. Jpn.}\ }\textbf {\bibinfo {volume} {76}},\ \bibinfo {pages} {034712} (\bibinfo {year} {2007})}\BibitemShut {NoStop}%
\bibitem [{\citenamefont {Fujimoto}\ and\ \citenamefont {Yip}(2012)}]{Fujimoto2012}%
  \BibitemOpen
  \bibfield  {author} {\bibinfo {author} {\bibfnamefont {S.}~\bibnamefont {Fujimoto}}\ and\ \bibinfo {author} {\bibfnamefont {S.~K.}\ \bibnamefont {Yip}},\ }\bibinfo {title} {{Aspects of Spintronics}},\ in\ \href {https://doi.org/10.1007/978-3-642-24624-1_8} {\emph {\bibinfo {booktitle} {Non-Centrosymmetric Superconductors: Introduction and Overview}}},\ \bibinfo {editor} {edited by\ \bibinfo {editor} {\bibfnamefont {E.}~\bibnamefont {Bauer}}\ and\ \bibinfo {editor} {\bibfnamefont {M.}~\bibnamefont {Sigrist}}}\ (\bibinfo  {publisher} {Springer Berlin Heidelberg},\ \bibinfo {address} {Berlin, Heidelberg},\ \bibinfo {year} {2012})\ pp.\ \bibinfo {pages} {247--266}\BibitemShut {NoStop}%
\bibitem [{\citenamefont {Miyamoto}\ \emph {et~al.}(2019)\citenamefont {Miyamoto}, \citenamefont {Sano}, \citenamefont {Miyakawa}, \citenamefont {Niinomi}, \citenamefont {Toyoda}, \citenamefont {Vall\'{e}s},\ and\ \citenamefont {Omatsu}}]{miy19}%
  \BibitemOpen
  \bibfield  {author} {\bibinfo {author} {\bibfnamefont {K.}~\bibnamefont {Miyamoto}}, \bibinfo {author} {\bibfnamefont {K.}~\bibnamefont {Sano}}, \bibinfo {author} {\bibfnamefont {T.}~\bibnamefont {Miyakawa}}, \bibinfo {author} {\bibfnamefont {H.}~\bibnamefont {Niinomi}}, \bibinfo {author} {\bibfnamefont {K.}~\bibnamefont {Toyoda}}, \bibinfo {author} {\bibfnamefont {A.}~\bibnamefont {Vall\'{e}s}},\ and\ \bibinfo {author} {\bibfnamefont {T.}~\bibnamefont {Omatsu}},\ }\bibfield  {title} {\bibinfo {title} {{Generation of high-quality terahertz OAM mode based on soft-aperture difference frequency generation}},\ }\href {https://doi.org/10.1364/OE.27.031840} {\bibfield  {journal} {\bibinfo  {journal} {Opt. Express}\ }\textbf {\bibinfo {volume} {27}},\ \bibinfo {pages} {31840} (\bibinfo {year} {2019})}\BibitemShut {NoStop}%
\bibitem [{\citenamefont {Arikawa}\ \emph {et~al.}(2020)\citenamefont {Arikawa}, \citenamefont {Hiraoka}, \citenamefont {Morimoto}, \citenamefont {Blanchard}, \citenamefont {Tani}, \citenamefont {Tanaka}, \citenamefont {Sakai}, \citenamefont {Kitajima}, \citenamefont {Sasaki},\ and\ \citenamefont {Tanaka}}]{ari20}%
  \BibitemOpen
  \bibfield  {author} {\bibinfo {author} {\bibfnamefont {T.}~\bibnamefont {Arikawa}}, \bibinfo {author} {\bibfnamefont {T.}~\bibnamefont {Hiraoka}}, \bibinfo {author} {\bibfnamefont {S.}~\bibnamefont {Morimoto}}, \bibinfo {author} {\bibfnamefont {F.}~\bibnamefont {Blanchard}}, \bibinfo {author} {\bibfnamefont {S.}~\bibnamefont {Tani}}, \bibinfo {author} {\bibfnamefont {T.}~\bibnamefont {Tanaka}}, \bibinfo {author} {\bibfnamefont {K.}~\bibnamefont {Sakai}}, \bibinfo {author} {\bibfnamefont {H.}~\bibnamefont {Kitajima}}, \bibinfo {author} {\bibfnamefont {K.}~\bibnamefont {Sasaki}},\ and\ \bibinfo {author} {\bibfnamefont {K.}~\bibnamefont {Tanaka}},\ }\bibfield  {title} {\bibinfo {title} {{Transfer of orbital angular momentum of light to plasmonic excitations in metamaterials}},\ }\href {https://doi.org/10.1126/sciadv.aay1977} {\bibfield  {journal} {\bibinfo  {journal} {Sci. Adv.}\ }\textbf {\bibinfo {volume} {6}},\ \bibinfo {pages} {eaay1977} (\bibinfo {year} {2020})}\BibitemShut {NoStop}%
\bibitem [{\citenamefont {Kohda}\ \emph {et~al.}(2012)\citenamefont {Kohda}, \citenamefont {Lechner}, \citenamefont {Kunihashi}, \citenamefont {Dollinger}, \citenamefont {Olbrich}, \citenamefont {Sch\"onhuber}, \citenamefont {Caspers}, \citenamefont {Bel'kov}, \citenamefont {Golub}, \citenamefont {Weiss}, \citenamefont {Richter}, \citenamefont {Nitta},\ and\ \citenamefont {Ganichev}}]{koh12}%
  \BibitemOpen
  \bibfield  {author} {\bibinfo {author} {\bibfnamefont {M.}~\bibnamefont {Kohda}}, \bibinfo {author} {\bibfnamefont {V.}~\bibnamefont {Lechner}}, \bibinfo {author} {\bibfnamefont {Y.}~\bibnamefont {Kunihashi}}, \bibinfo {author} {\bibfnamefont {T.}~\bibnamefont {Dollinger}}, \bibinfo {author} {\bibfnamefont {P.}~\bibnamefont {Olbrich}}, \bibinfo {author} {\bibfnamefont {C.}~\bibnamefont {Sch\"onhuber}}, \bibinfo {author} {\bibfnamefont {I.}~\bibnamefont {Caspers}}, \bibinfo {author} {\bibfnamefont {V.~V.}\ \bibnamefont {Bel'kov}}, \bibinfo {author} {\bibfnamefont {L.~E.}\ \bibnamefont {Golub}}, \bibinfo {author} {\bibfnamefont {D.}~\bibnamefont {Weiss}}, \bibinfo {author} {\bibfnamefont {K.}~\bibnamefont {Richter}}, \bibinfo {author} {\bibfnamefont {J.}~\bibnamefont {Nitta}},\ and\ \bibinfo {author} {\bibfnamefont {S.~D.}\ \bibnamefont {Ganichev}},\ }\bibfield  {title} {\bibinfo {title} {{Gate-controlled persistent spin helix state in (In,Ga)As quantum wells}},\ }\href
  {https://doi.org/10.1103/PhysRevB.86.081306} {\bibfield  {journal} {\bibinfo  {journal} {Phys. Rev. B}\ }\textbf {\bibinfo {volume} {86}},\ \bibinfo {pages} {081306} (\bibinfo {year} {2012})}\BibitemShut {NoStop}%
\bibitem [{\citenamefont {Bernevig}\ \emph {et~al.}(2006)\citenamefont {Bernevig}, \citenamefont {Orenstein},\ and\ \citenamefont {Zhang}}]{ber06}%
  \BibitemOpen
  \bibfield  {author} {\bibinfo {author} {\bibfnamefont {B.~A.}\ \bibnamefont {Bernevig}}, \bibinfo {author} {\bibfnamefont {J.}~\bibnamefont {Orenstein}},\ and\ \bibinfo {author} {\bibfnamefont {S.-C.}\ \bibnamefont {Zhang}},\ }\bibfield  {title} {\bibinfo {title} {{Exact SU(2) Symmetry and Persistent Spin Helix in a Spin-Orbit Coupled System}},\ }\href {https://doi.org/10.1103/PhysRevLett.97.236601} {\bibfield  {journal} {\bibinfo  {journal} {Phys. Rev. Lett.}\ }\textbf {\bibinfo {volume} {97}},\ \bibinfo {pages} {236601} (\bibinfo {year} {2006})}\BibitemShut {NoStop}%
\bibitem [{\citenamefont {Hall}(1996)}]{hall_1996}%
  \BibitemOpen
  \bibfield  {author} {\bibinfo {author} {\bibfnamefont {D.~G.}\ \bibnamefont {Hall}},\ }\bibfield  {title} {\bibinfo {title} {{Vector-beam solutions of Maxwell’s wave equation}},\ }\href {https://doi.org/10.1364/OL.21.000009} {\bibfield  {journal} {\bibinfo  {journal} {Opt. Lett.}\ }\textbf {\bibinfo {volume} {21}},\ \bibinfo {pages} {9} (\bibinfo {year} {1996})}\BibitemShut {NoStop}%
\bibitem [{\citenamefont {Miller}\ \emph {et~al.}(2003)\citenamefont {Miller}, \citenamefont {Zumb\"uhl}, \citenamefont {Marcus}, \citenamefont {Lyanda-Geller}, \citenamefont {Goldhaber-Gordon}, \citenamefont {Campman},\ and\ \citenamefont {Gossard}}]{mil03}%
  \BibitemOpen
  \bibfield  {author} {\bibinfo {author} {\bibfnamefont {J.~B.}\ \bibnamefont {Miller}}, \bibinfo {author} {\bibfnamefont {D.~M.}\ \bibnamefont {Zumb\"uhl}}, \bibinfo {author} {\bibfnamefont {C.~M.}\ \bibnamefont {Marcus}}, \bibinfo {author} {\bibfnamefont {Y.~B.}\ \bibnamefont {Lyanda-Geller}}, \bibinfo {author} {\bibfnamefont {D.}~\bibnamefont {Goldhaber-Gordon}}, \bibinfo {author} {\bibfnamefont {K.}~\bibnamefont {Campman}},\ and\ \bibinfo {author} {\bibfnamefont {A.~C.}\ \bibnamefont {Gossard}},\ }\bibfield  {title} {\bibinfo {title} {{Gate-Controlled Spin-Orbit Quantum Interference Effects in Lateral Transport}},\ }\href {https://doi.org/10.1103/PhysRevLett.90.076807} {\bibfield  {journal} {\bibinfo  {journal} {Phys. Rev. Lett.}\ }\textbf {\bibinfo {volume} {90}},\ \bibinfo {pages} {076807} (\bibinfo {year} {2003})}\BibitemShut {NoStop}%
\bibitem [{\citenamefont {Yama}\ \emph {et~al.}(2021)\citenamefont {Yama}, \citenamefont {Tatsuno}, \citenamefont {Kato},\ and\ \citenamefont {Matsuo}}]{yam21}%
  \BibitemOpen
  \bibfield  {author} {\bibinfo {author} {\bibfnamefont {M.}~\bibnamefont {Yama}}, \bibinfo {author} {\bibfnamefont {M.}~\bibnamefont {Tatsuno}}, \bibinfo {author} {\bibfnamefont {T.}~\bibnamefont {Kato}},\ and\ \bibinfo {author} {\bibfnamefont {M.}~\bibnamefont {Matsuo}},\ }\bibfield  {title} {\bibinfo {title} {{Spin pumping of two-dimensional electron gas with Rashba and Dresselhaus spin-orbit interactions}},\ }\href {https://doi.org/10.1103/PhysRevB.104.054410} {\bibfield  {journal} {\bibinfo  {journal} {Phys. Rev. B}\ }\textbf {\bibinfo {volume} {104}},\ \bibinfo {pages} {054410} (\bibinfo {year} {2021})}\BibitemShut {NoStop}%
\bibitem [{\citenamefont {Yama}\ \emph {et~al.}(2023)\citenamefont {Yama}, \citenamefont {Matsuo},\ and\ \citenamefont {Kato}}]{yam23}%
  \BibitemOpen
  \bibfield  {author} {\bibinfo {author} {\bibfnamefont {M.}~\bibnamefont {Yama}}, \bibinfo {author} {\bibfnamefont {M.}~\bibnamefont {Matsuo}},\ and\ \bibinfo {author} {\bibfnamefont {T.}~\bibnamefont {Kato}},\ }\bibfield  {title} {\bibinfo {title} {{Effect of vertex corrections on the enhancement of Gilbert damping in spin pumping into a two-dimensional electron gas}},\ }\href {https://doi.org/10.1103/PhysRevB.107.174414} {\bibfield  {journal} {\bibinfo  {journal} {Phys. Rev. B}\ }\textbf {\bibinfo {volume} {107}},\ \bibinfo {pages} {174414} (\bibinfo {year} {2023})}\BibitemShut {NoStop}%
\bibitem [{\citenamefont {Masselink}\ \emph {et~al.}(1985)\citenamefont {Masselink}, \citenamefont {Pearah}, \citenamefont {Klem}, \citenamefont {Peng}, \citenamefont {Morko\ifmmode~\mbox{\c{c}}\else \c{c}\fi{}}, \citenamefont {Sanders},\ and\ \citenamefont {Chang}}]{mas85}%
  \BibitemOpen
  \bibfield  {author} {\bibinfo {author} {\bibfnamefont {W.~T.}\ \bibnamefont {Masselink}}, \bibinfo {author} {\bibfnamefont {P.~J.}\ \bibnamefont {Pearah}}, \bibinfo {author} {\bibfnamefont {J.}~\bibnamefont {Klem}}, \bibinfo {author} {\bibfnamefont {C.~K.}\ \bibnamefont {Peng}}, \bibinfo {author} {\bibfnamefont {H.}~\bibnamefont {Morko\ifmmode~\mbox{\c{c}}\else \c{c}\fi{}}}, \bibinfo {author} {\bibfnamefont {G.~D.}\ \bibnamefont {Sanders}},\ and\ \bibinfo {author} {\bibfnamefont {Y.-C.}\ \bibnamefont {Chang}},\ }\bibfield  {title} {\bibinfo {title} {{Absorption coefficients and exciton oscillator strengths in AlGaAs-GaAs superlattices}},\ }\href {https://doi.org/10.1103/PhysRevB.32.8027} {\bibfield  {journal} {\bibinfo  {journal} {Phys. Rev. B}\ }\textbf {\bibinfo {volume} {32}},\ \bibinfo {pages} {8027} (\bibinfo {year} {1985})}\BibitemShut {NoStop}%
\bibitem [{SM()}]{SM}%
  \BibitemOpen
  \href@noop {} {}\bibinfo {note} {See Supplement Material at [url] for GIF files demonstrating the dynamics of local spins in 2DEGs induced by pulsed vortex beams.}\BibitemShut {Stop}%
\bibitem [{\citenamefont {Dey}\ \emph {et~al.}(2025)\citenamefont {Dey}, \citenamefont {Nandy},\ and\ \citenamefont {Saha}}]{dey25}%
  \BibitemOpen
  \bibfield  {author} {\bibinfo {author} {\bibfnamefont {A.}~\bibnamefont {Dey}}, \bibinfo {author} {\bibfnamefont {A.~K.}\ \bibnamefont {Nandy}},\ and\ \bibinfo {author} {\bibfnamefont {K.}~\bibnamefont {Saha}},\ }\bibfield  {title} {\bibinfo {title} {{Current-induced spin polarisation in Rashba–Dresselhaus systems under different point groups}},\ }\href {https://doi.org/10.1088/1367-2630/adac88} {\bibfield  {journal} {\bibinfo  {journal} {New J. Phys.}\ }\textbf {\bibinfo {volume} {27}},\ \bibinfo {pages} {013024} (\bibinfo {year} {2025})}\BibitemShut {NoStop}%
\bibitem [{\citenamefont {Qiu}\ and\ \citenamefont {Bader}(1999)}]{qiu99}%
  \BibitemOpen
  \bibfield  {author} {\bibinfo {author} {\bibfnamefont {Z.}~\bibnamefont {Qiu}}\ and\ \bibinfo {author} {\bibfnamefont {S.}~\bibnamefont {Bader}},\ }\bibfield  {title} {\bibinfo {title} {{Surface magneto-optic Kerr effect (SMOKE)}},\ }\href {https://doi.org/https://doi.org/10.1016/S0304-8853(99)00311-X} {\bibfield  {journal} {\bibinfo  {journal} {J. Magn. Magn. Mater.}\ }\textbf {\bibinfo {volume} {200}},\ \bibinfo {pages} {664} (\bibinfo {year} {1999})}\BibitemShut {NoStop}%
\bibitem [{\citenamefont {Qiu}\ and\ \citenamefont {Bader}(2000)}]{qiu00}%
  \BibitemOpen
  \bibfield  {author} {\bibinfo {author} {\bibfnamefont {Z.~Q.}\ \bibnamefont {Qiu}}\ and\ \bibinfo {author} {\bibfnamefont {S.~D.}\ \bibnamefont {Bader}},\ }\bibfield  {title} {\bibinfo {title} {{Surface magneto-optic Kerr effect}},\ }\href {https://doi.org/10.1063/1.1150496} {\bibfield  {journal} {\bibinfo  {journal} {Rev. Sci. Instrum.}\ }\textbf {\bibinfo {volume} {71}},\ \bibinfo {pages} {1243} (\bibinfo {year} {2000})}\BibitemShut {NoStop}%
\bibitem [{\citenamefont {Zhou}\ \emph {et~al.}(2020)\citenamefont {Zhou}, \citenamefont {Vernier}, \citenamefont {Agnus}, \citenamefont {Eimer}, \citenamefont {Lin},\ and\ \citenamefont {Zhai}}]{zhou}%
  \BibitemOpen
  \bibfield  {author} {\bibinfo {author} {\bibfnamefont {X.}~\bibnamefont {Zhou}}, \bibinfo {author} {\bibfnamefont {N.}~\bibnamefont {Vernier}}, \bibinfo {author} {\bibfnamefont {G.}~\bibnamefont {Agnus}}, \bibinfo {author} {\bibfnamefont {S.}~\bibnamefont {Eimer}}, \bibinfo {author} {\bibfnamefont {W.}~\bibnamefont {Lin}},\ and\ \bibinfo {author} {\bibfnamefont {Y.}~\bibnamefont {Zhai}},\ }\bibfield  {title} {\bibinfo {title} {{Highly Anisotropic Magnetic Domain Wall Behavior in In-Plane Magnetic Films}},\ }\href {https://doi.org/10.1103/PhysRevLett.125.237203} {\bibfield  {journal} {\bibinfo  {journal} {Phys. Rev. Lett.}\ }\textbf {\bibinfo {volume} {125}},\ \bibinfo {pages} {237203} (\bibinfo {year} {2020})}\BibitemShut {NoStop}%
\end{thebibliography}%

\end{document}